\documentclass[10pt,twocolumn]{IEEEtran}

\usepackage{amssymb,amsmath,color,graphicx, amsbsy}
\usepackage{graphicx}
\usepackage{subfigure}
\usepackage{cite}
\usepackage{algorithm}
\usepackage{amsmath}
\usepackage{amsfonts}
\usepackage{amssymb}
\usepackage{graphicx}
\usepackage{epstopdf}
\usepackage{lettrine}
\usepackage{xurl}
\usepackage{algpseudocode} 
\usepackage{tikz}
\usepackage{hyperref}
\usepackage{multirow}
\newcommand*\circled[1]{\tikz[baseline=(char.base)]{
    \node[shape=circle, draw, inner sep=1pt] (char) {#1};}}
    
\usepackage{multirow}
\usepackage{colortbl}
\usepackage{etoolbox}

\begin{document}

\title{A Data-Driven
Convergence Bidding Strategy \\ Based on Reverse Engineering of Market Participants' Performance: A Case of California ISO}

\author{Ehsan Samani,~\IEEEmembership{Student Member, IEEE}, Mahdi Kohansal, \IEEEmembership{Member,~IEEE}, \\
and Hamed Mohsenian-Rad,~\IEEEmembership{Fellow, IEEE}
\thanks{The authors are with the Department of Electrical and Computer Engineering, University of California, Riverside, CA, USA, 92521. This work was supported by the National Science Foundation (NSF) grant 1711944. The corresponding author is Hamed Mohsenian-Rad; e-mail: hamed@ece.ucr.edu.}\vspace{-0.3cm}}

\maketitle

\thispagestyle{empty}

\pagestyle{empty}

\begin{abstract}

Convergence bidding, a.k.a., virtual bidding, has been widely adopted in wholesale electricity markets in recent years. 
It provides opportunities for market participants to arbitrage on the \emph{difference} between the day-ahead market locational marginal prices  and the real-time market locational marginal prices. 
Given the fact that convergence bids (CBs) have a significant impact on the operation of electricity markets, it is important to understand \emph{how} market participants \emph{strategically} select their CBs in \emph{real-world} electricity markets.
We address this open problem with focus on the electricity market that is operated by the California Independent System Operator (ISO). In this regard, we use the publicly available electricity market data \emph{to learn, characterize, and evaluate} different types of convergence bidding \emph{strategies} that are currently used by market participants.
Our analysis includes developing a \emph{data-driven reverse engineering} method that we apply to three years of real-world California ISO market data. Our analysis involves feature selection and density-based data clustering. It results in identifying \emph{three main clusters} of CB strategies in the California ISO market. 
Different characteristics and the performance of each cluster of strategies are analyzed.
Interestingly, we unmask a common real-world strategy that does \emph{not} match any of the existing strategic convergence bidding methods in the literature. 
Next, we build upon the lessons learned from the advantages and disadvantages of the existing real-world strategies in order to propose a new CB strategy that can significantly \emph{outperform} them.
Our analysis includes developing a new strategy for convergence bidding.
The new strategy has three steps: net profit maximization by capturing price spikes, dynamic node labeling, and strategy selection algorithm.
We show through case studies that the annual net profit for the most lucrative market participants can increase by over 40\% if the proposed convergence bidding strategy is used.

\vspace{0.25cm}


\end{abstract}

\section*{Nomenclature}


\noindent \textbf{Abbreviations}

\vspace{+0.2cm}

\noindent \begin{tabular}{ l p{6.55cm} }

CB &  Convergence Bid \\

DAM & Day-Ahead  Market \\

RTM  & Real-Time  Market \\

LMP &  Locational  Marginal  Price \\

D-LMP &  Day-Ahead Market LMP \\

R-LMP &  Real-Time Market LMP \\

ISO & Independent System Operator \\

Pnode & Pricing Node \\

\end{tabular}

\vspace{+0.4cm}

\noindent \begin{tabular}{ l p{6.55cm} }

APnode & Aggregated Pricing Node \\

DLAP & Default Load Aggregated Point \\

MILP & Mixed-Integer Linear
Programming \\

HDBSCAN & Hierarchical Density-Based Spatial Clustering of Applications with Noise \\

\end{tabular}

\vspace{+0.2cm}

\noindent \textbf{Indices, Sets, and Symbols}

\vspace{+0.2cm}

\noindent \begin{tabular}{ l p{6.55cm} }

$T$ & Set of all the time intervals \\

$t$ & Index of time interval \\

$ (\boldsymbol{\cdot})^{*} $ & Symbol for average value \\

$ (\boldsymbol{\cdot})^{min/max} $ & Symbol for upper/lower limit \\ 

\end{tabular}

\vspace{+0.2cm}

\noindent \textbf{Parameters}

\vspace{+0.2cm}

\noindent \begin{tabular}{ l p{6.55cm} }

$\Delta$ & The distance of price bid from the average hourly D-LMP \\

$\lambda$ & D-LMP \\

$\pi$ & R-LMP \\

$\delta$ & Difference between D-LMP and R-LMP \\

$\epsilon$ & Relatively small number \\

$M$ & Sufficiently large number \\

$\theta$ & Objective function threshold in Algorithm 1 \\

a & Accuracy of forecasting \\

\end{tabular}

\vspace{+0.2cm}

\noindent \textbf{Variables}

\vspace{+0.2cm}

\noindent \begin{tabular}{ l p{6.55cm} }

$\eta$ & Net profit \\

$L$ & Loss / negative profit \\

$P$ & Profit / positive profit \\

$x$ & Submitted price bid \\

$m$ & Distance from the average D-LMP captured as a spike by an optimal CB \\

$b^1$, $b^2$ & Auxiliary binary variables \\

$z$ & Auxiliary continuous variable \\

\end{tabular}

\section{Introduction} \label{sec:introduction}
\subsection{Background: Convergence Bidding} 
\begin{figure}[t]
 \centering
{\scalebox{0.66}{\includegraphics*{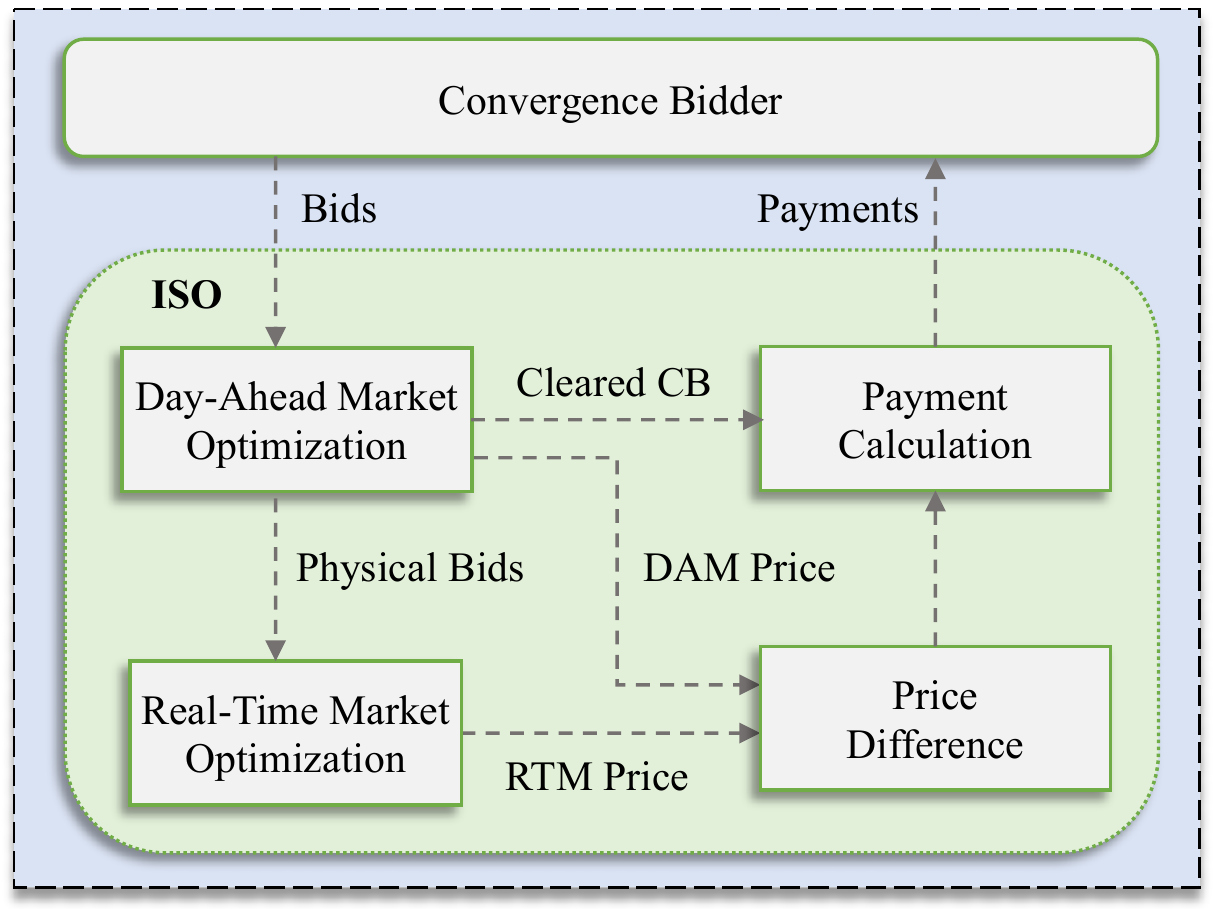}}}
\caption{The process of clearing CBs in wholesale electricity markets \cite{samani2020cb}.}
\label{fig:out1}
\end{figure}

Convergence bidding, a.k.a., virtual bidding, is a market mechanism that is used by Independent System Operators (ISOs) in two-settlement wholesale electricity markets to reduce the gap between the day-ahead market (DAM) prices and the real-time market (RTM) prices in order to increase market efficiency \cite{hogan2016,2015Financial0MIT}. 
A supply convergence bid (CB) is a bid to sell energy in DAM and buy the \emph{same amount} of energy in RTM. A demand CB is a bid to buy energy in DAM and sell the \emph{same amount} of energy in RTM \cite{samani2020cb}. 
While CBs are virtual, i.e., only financial and not physical, they are cleared in DAM together with physical supply and demand bids.
If a supply CB is cleared in DAM, then the bidder is credited at the DAM price and charged at the RTM price; and if a demand CB is cleared in DAM, then the bidder is charged at the DAM price and credited at the RTM price. In both cases, the \emph{difference} between the earning or loss is paid to the convergence bidder.
The process of clearing CBs and the related payment calculation is outlined in Fig. \ref{fig:out1}. The payment is calculated by multiplying the cleared amount of energy by the \emph{difference} between the DAM locational marginal price (D-LMP) and the RTM locational marginal price (R-LMP). 

Most ISOs in the United States, including the California ISO, have already adopted CBs \cite{Ref_CAISO_CB_Document, 2015Virtual0Pjm}. CBs are currently understood to play a critical role in electricity markets, e.g., to improve market efficiency, reduce the price gap between D-LMP and R-LMP, and help with the integration of renewable energy resources, e.g., see the various studies in \cite{kazempour2017value,kazempour2017value2,woo2015virtual,hogan2006revenue}. 
%

\color{black}
In this paper, we seek to answer a series of research questions related to CB strategies:
1) How many market participants submit CBs in the California ISO market, and what are the characteristics of their CBs?
2) How do market participants shape their CB strategy, in particular with respect to the choice of their price bids? 
3) How does the reality of the CB strategies in the California ISO market match the existing literature in this field?
4) What are the most common strategies that are used by the CB market participants in the California ISO market? 
5) Is it possible to learn from the current CB strategies in the California ISO market and propose a new strategy that can significantly outperform them?
6) Could a CB strategy that is seemingly unprofitable comprise part of an enhanced new composite bidding strategy?
\color{black}
For the rest of this paper, we will refer to the above research questions as Research Question 1 to Research Question 6, respectively. We will address and refer to these questions throughout the paper. \color{black}


\subsection{Summary of Contributions and Discoveries} \label{sec:intr:cntr}
While the basic principles of convergence bidding are studied in the academic literature and industry reports, there is currently a gap in this field about understanding the strategy and behavior of CB market participants in real-world electricity markets.
This is a critical subject because the way that market participants select their CBs can ultimately shape the impact of CBs on electricity markets. Addressing this open problem is the focus of this paper. Accordingly, the main discoveries and contributions in this paper are  as follows:

\begin{itemize}
\vspace{0.1cm}

\item Three years of real-world market data from the California ISO market are investigated to understand the behavior of CB market participants. The analysis is comprehensive; it looks into all the submitted CBs, D-LMPs, R-LMPs, and the net cleared CBs. 
\color{black}
The convergence bidders that are most present in the market are identified based on different metrics; and their CBs are analyzed \color{black} in terms of the number of submitted CBs, the number of participated locations, the type of submitted CBs, the number of steps for the submitted CBs, and the quantity of 
in MWh.



\vspace{0.1cm}

\item The \emph{features} for the strategy of the submitted CBs are extracted; and by using a density-based clustering algorithm, \emph{three main clusters} of CB strategies are identified.
The characteristics and the performance of each identified cluster of strategies are analyzed and some of their \emph{advantages} and \emph{disadvantages} are investigated. 
Next, the identified strategies are \emph{reverse engineered}, i.e., their key steps are identified such that we can implement them for a market participant. 
The purpose of this reverse engineering task is two-fold. First, it can shed light on \emph{how} CB market participants behave. This by itself is an important study and the results can be insightful to ISOs and policy makers. Second, it serves as means for us to develop a \emph{new} and better convergence bidding strategy based on what we learn from the current state of practice. 

\vspace{0.1cm}

\item Our analysis also unmasks two interesting discoveries. First, one of the most common real-world CB strategies in the California ISO market does \emph{not} match any of the strategic convergence bidding methods that currently exist in the research literature. Second, most of the exciting papers in the research literature are focused on one of the CB strategies that is \emph{less} common in practice among the CB market participants in the California ISO market. 

\vspace{0.1cm}

\item A new comprehensive convergence bidding strategy
is proposed to utilize the identified reverse engineered strategies based on their advantages and disadvantages under various market conditions.
To the best of our knowledge, this is the first composite CB strategy that is proposed in the literature. It is also the first CB strategy that is obtained by reverse engineering of existing real-world CB strategies.
The proposed strategy comprises \emph{three steps}: net profit maximization by capturing price spikes, dynamic node labeling, and strategy selection.
We show that the annual profit for the most lucrative market participant in the California ISO market can increase by 43\%; if the proposed bidding strategy is used. 

\vspace{0.1cm}
\end{itemize}

\subsection{Literature Review} \label{sec:intr:lit}
Despite the fact that CBs are widely adopted by ISOs in recent years, the current literature is still limited when it comes to the analysis of convergence bidding strategies. 

Some of the related papers include \cite{baltaoglu2018algorithmic,xiao2018risk,kohansal2020strategic,xiao2019risk,mehdipourpicha2020risk,xiao190optimal,wang2019machine}. 
In \cite{baltaoglu2018algorithmic}, an online learning algorithm is proposed to maximize the cumulative payoff over a finite number of CB trading sessions.
\color{black}
However, there is no discussion on how the proposed strategy is similar to or different from the strategies that are 
%
currently used by the market participants in practice. 
\color{black}
In \cite{xiao2018risk}, a stochastic optimization model is proposed to place CBs under different risk management scenarios.
\color{black}
The focus is on self-scheduling bids; therefore, the choice of the price components for the CBs is inherently not part of the analysis.  
\color{black}
In \cite{kohansal2020strategic}, a bi-level CB optimization problem is proposed, where the upper-level problem aims to maximize the profit for the convergence bidder and the lower-level is the economic dispatch problem. The authors in \cite{xiao190optimal} also proposed a bi-level stochastic optimization model for joint physical demand bidding and convergence bidding, for a strategic retailer in the short-term electricity market.
\color{black}
While the use of bi-level optimization  is insightful, it may not match the information available to CB market participants in practice. In fact, in practice, market participants do not have access to the detailed formulation of the economic dispatch problem that is solved by the ISO. They also do not have access to the comprehensive market data that are needed to solve the economic dispatch problem.  
\color{black} In summary, while the above papers do propose new convergence bidding strategies, they are not concerned with convergence bidding strategies that already exist in practice in the real-world convergence bidding markets. This is indeed still an open issue that needs to be explored, which we seek to address in this paper.  \color{black}
\color{black}

There are fundamental differences between the work in this paper and the few available studies related to the convergence bidding strategies that we listed in the previous paragraph. %
\emph{First}, none of the available works in the literature has proposed a data-driven convergence bidding strategy based on the real-world market data. 
\color{black}
\emph{Second}, some of the existing convergence bidding papers in the literature focus only on selecting the quantity (in MWh) of the bid, and not the price of the bid; i.e., they focus on self-scheduling strategies.
\color{black}
\emph{Third}, some of the papers in the literature tried to solve the market clearing process as part of their problem in order to consider the price component of CBs and to be able to use the resultant LMPs in the profit maximization problem.
However, as we know, the market clearing problem is a very complex optimization problem with many steps. Therefore, the authors in those papers had no choice but to significantly simplify the market model in order to solve their formulated optimization problems.
On the contrary, in this work, we focus on real-world market data and we analyze and reverse engineer the strategies of each convergence bidder in the California ISO market. 

Compared to the preliminary conference version of this work in \cite{Ehsan_ISGT}, the current journal submission has several new and important contributions. In fact, the two major tasks of reverse engineering the existing CB strategies as well as designing a new comprehensive convergence bidding strategy are both new in this journal version. This journal version also includes new results with respect to identifying the advantages and disadvantages of various real-world convergence bidding strategies in the California ISO market.

Finally, there is also a rich body of literature that studies CBs; but they are \emph{not} about understanding the strategies of CB market participants. In particular, there are papers that study the impact of CBs on electricity markets \cite{samani2020cb,2016Dynamic0Theis,oren2015,nyc2007,tang2016model,hadsell2007one,wolak2013,jha2019can,mather2017virtual,larrieu2015impact,tangimpact,you2019role,long2020exploring}. 
In \cite{samani2020cb}, a method is proposed to identify under what theoretical conditions a CB results in price divergence, instead of price convergence.
%
%
The impact of convergence bidding on the efficiency of the California ISO market and impact on price convergence is studied in \cite{oren2015} and also in \cite{jha2019can}.
%
There are also a few papers in the literature that are concerned with the potential to manipulate the wholesale electricity market by using CBs \cite{ftr2013,AnotherFTR,birge2018limits,shan2017simulation,choi2016economic,tajer2017false,goodbad2010}.
In \cite{AnotherFTR}, an equilibrium model is developed to study the cross-product manipulation in financial transmission right and two-settlement energy markets.
The concept of cyber attacks in wholesale electricity markets with virtual bidding activities is analyzed in \cite{choi2016economic}. A framework is proposed to evaluate the economic profit of an attacker who conducts a topology data attack using CBs.

\color{black}
In the second part of this paper in Section IV, we look at the strategic bidding problem from the viewpoint of the market participants. In this regard, the viewpoint in this paper is similar to those in \cite{baltaoglu2018algorithmic,7307233,6767152,7038219,7892020,7524679,7533471}; all of which discuss developing new bidding strategies in electricity markets.
\color{black}

\section{Overview of the Real-World CB Market Participation Data in the California ISO} \label{sec:gnral}

In this section, we provide an overview of the CB market participation in the California ISO electricity market based on the real-world market data. All the raw data in this study are available in \cite{OASIScaiso}. The analysis in this section
\color{black} will address Research Question 1. It will also \color{black} set the stage for the data-driven reverse engineering work in Section III.

\subsection{Analyzing the Market Data} \label{sec:gnral:data}
Three years of market data from the California ISO electricity market, during 2017, 2018, and 2019, are analyzed.

\color{black}
A CB that is submitted to the California ISO electricity market must contain four pieces of information as follows: i) step-wise quantities (MWh), ii) step-wise prices\footnote{\color{black} Throughout this paper, we refer to the price bids, which are expressed in \$/MWh, as the \emph{price components} or the \emph{price values} of the convergence bids.} (\$/MWh), iii) the type of the CB, which can be either a demand CB or a supply CB, and iv) the nodal location of the CB.
In the California ISO electricity market, the CB market participants can submit up to ten steps of quantity and price pairs in each bid. 
It should be mentioned that, throughout this paper, if a CB is multi-step, then the maximum quantity of the different steps of the same submitted CB is considered as its \emph{quantity}.


\color{black}

In this study, we focused on the aggregated pricing nodes (APnodes) in the California ISO market.
\color{black}
As defined by the California ISO, an APnode is a trading hub, a load aggregation point, or any group of multiple pricing nodes (Pnodes)\color{black}\cite{CAISOBPMDAv19}.
\color{black}
The reason that APnodes are the focus of this study is that, most of the submitted CBs in the California ISO market are at the APnodes. There is no practical advantage to look into any higher locational resolution beyond APnodes. With over two thousand APnodes across the state of California, focusing on the APnodes in this study already required handling a huge amount of real-world market data.
\color{black}
Accordingly, we examined a total of 2265 APnodes; out of which a total of 475 APnodes hosted at least one CB at any time during the three-year period of this study. On average, a total of 387 APnodes hosted at least one CB during each month. 

\begin{figure}[t]
\centering
{\scalebox{0.45}{\includegraphics*{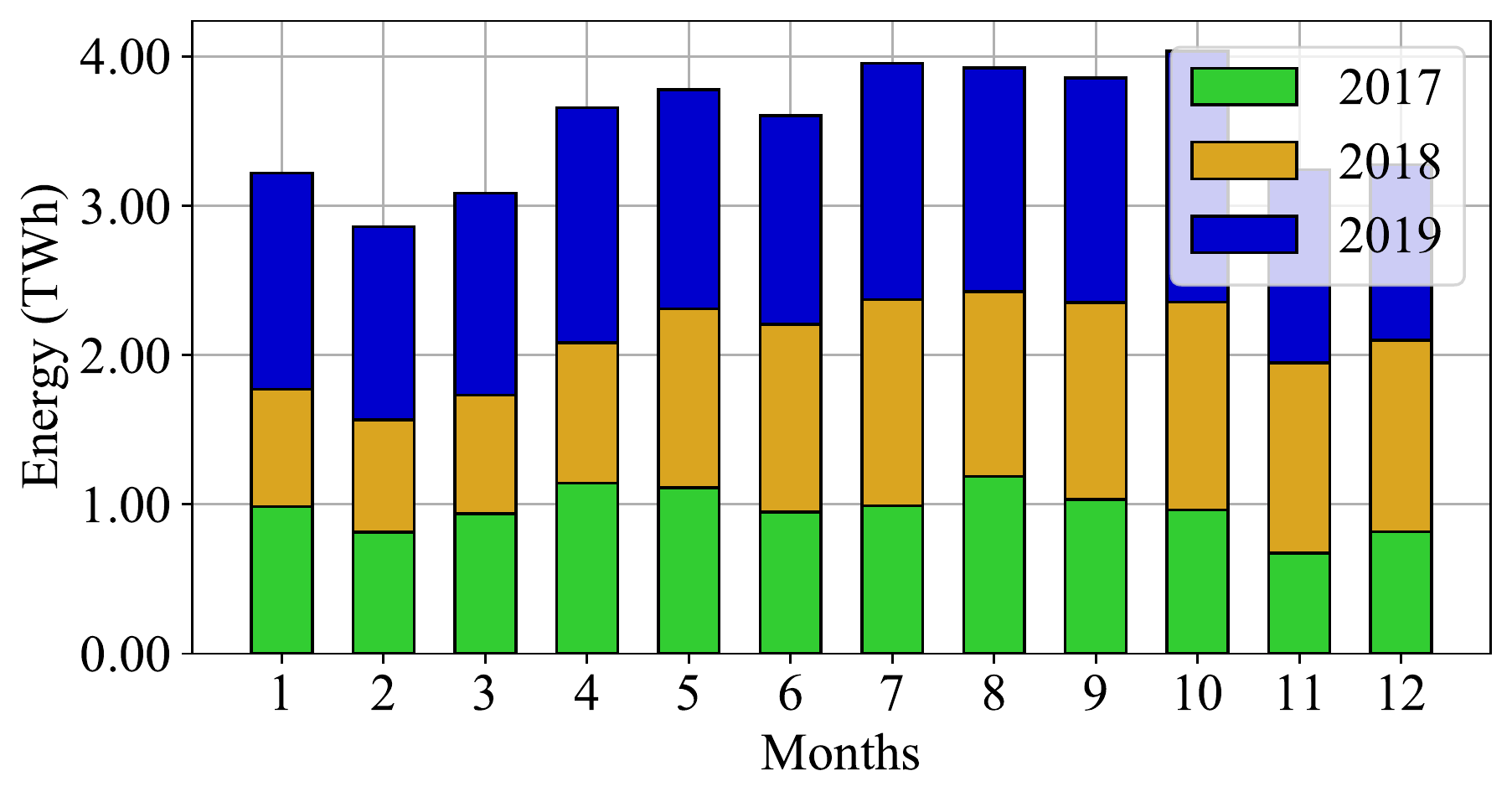}}} 
\caption{Total monthly amount of cleared energy by CBs for each year.}
\label{fig:mon_energy}
\end{figure}

\begin{figure}[t]
\centering
{\scalebox{0.45}{\includegraphics*{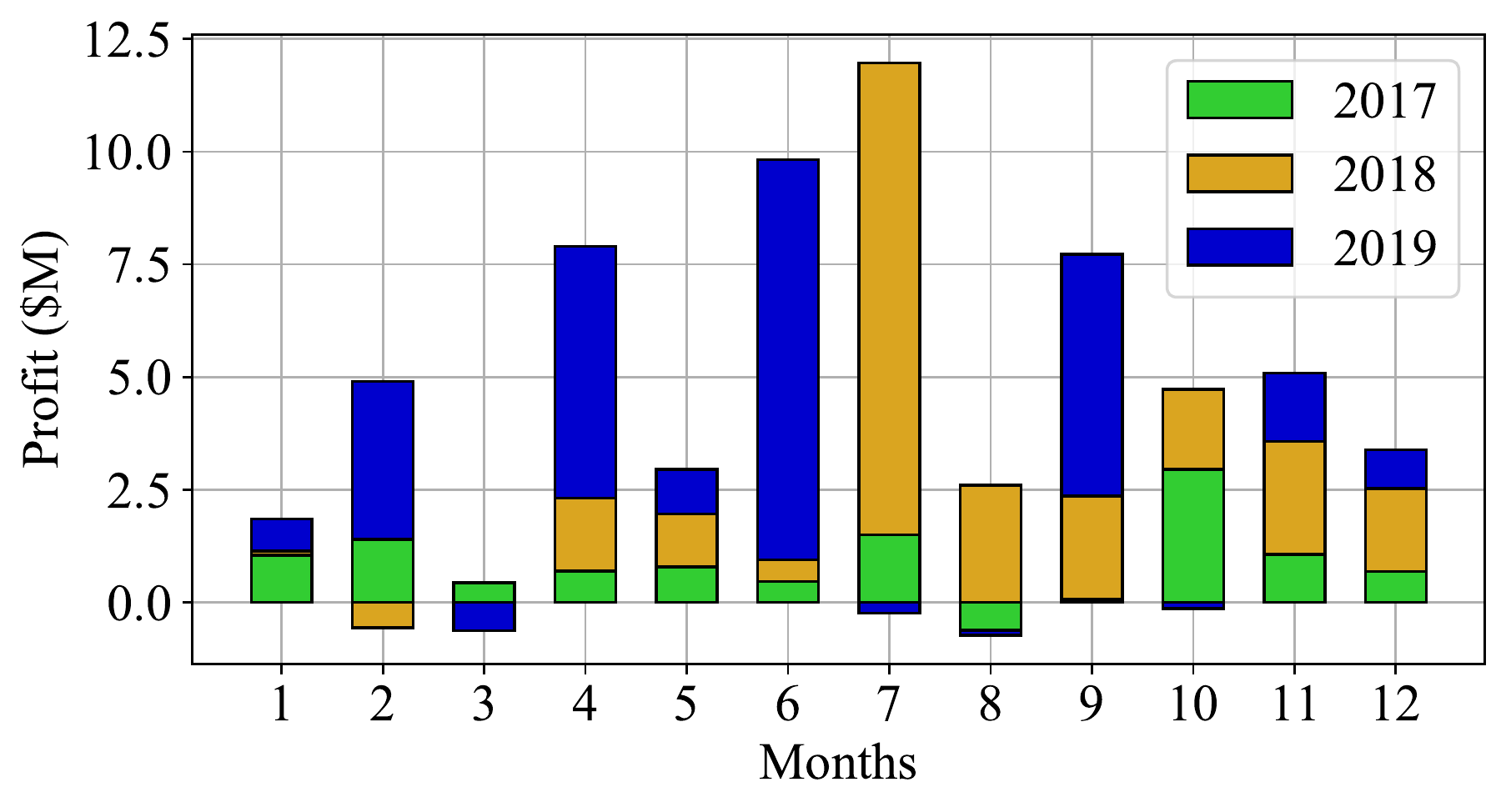}}}
\caption{Total monthly net profit by convergence bidders for each year.}
\label{fig:mon_prof}
\end{figure}

\begin{table}[t]
\color{black}
\centering
\caption {Selection of the CB Market Participants with Considerable Presence based on the Four Introduced Metrics across a Total of 101 Convergence Bidders.}
 \label{tab1}
\begin{center}
   \begin{tabular}{| c | c | c | c | c |}
   \hline

Alias ID & \circled{1} & \circled{2}  & \circled{3} & \circled{4} \\ \hline
1 & \cellcolor{gray!25}20.97 & 1.49  & \cellcolor{gray!25}11.06 & 1.93 \\ \hline
2 & \cellcolor{gray!25}8.53  & \cellcolor{gray!25}12.82 & \cellcolor{gray!25}5.61  & \cellcolor{gray!25}7.00 \\ \hline
3 & \cellcolor{gray!25}6.54  & \cellcolor{gray!25}8.92  & 1.68  & 2.51 \\ \hline
4 & \cellcolor{gray!25}6.05  & \cellcolor{gray!25}13.13 & \cellcolor{gray!25}3.00  & \cellcolor{gray!25}6.96 \\ \hline
5 & \cellcolor{gray!25}6.02  & 1.26  & \cellcolor{gray!25}16.26 & \cellcolor{gray!25}3.47 \\ \hline
6 & \cellcolor{gray!25}5.63  & 1.50  & \cellcolor{gray!25}7.09  & \cellcolor{gray!25}3.55 \\ \hline
7 & \cellcolor{gray!25}5.26  & 1.16  & 2.28  & 0.50 \\ \hline
8 & \cellcolor{gray!25}3.87  & 1.11  & \cellcolor{gray!25}2.34  & 1.21 \\ \hline
9 & \cellcolor{gray!25}3.69  & \cellcolor{gray!25}7.15  & \cellcolor{gray!25}10.56 & \cellcolor{gray!25}17.56 \\ \hline
10 & \cellcolor{gray!25}2.56  & \cellcolor{gray!25}3.30  & 0.68  & 1.06 \\ \hline
11 & 2.49  & \cellcolor{gray!25}5.62  & 0.31  & 0.72 \\ \hline
12 & 2.31  & \cellcolor{gray!25}2.92  & 1.11  & 1.47 \\ \hline
13 & 1.91  & 2.24  & \cellcolor{gray!25}5.77  & \cellcolor{gray!25}5.82 \\ \hline
14 & 1.67  & \cellcolor{gray!25}3.63  & 0.28  & 0.62 \\ \hline
15 & 1.58  & \cellcolor{gray!25}3.53  & 0.27  & 0.62 \\ \hline
16 & 1.36  & \cellcolor{gray!25}2.37  & 1.76  & 2.10 \\ \hline
17 & 1.21  & 1.90  & \cellcolor{gray!25}4.58  & \cellcolor{gray!25}5.10 \\ \hline
18 & 1.11  & 1.12  & 2.20  & \cellcolor{gray!25}2.71 \\ \hline
19 & 1.04  & 1.94  & \cellcolor{gray!25}2.50  & \cellcolor{gray!25}3.69 \\ \hline
20 & 0.19  & 0.41  & 1.88  & \cellcolor{gray!25}4.30 \\ \hline
Total & 84.0\% & 77.5\% & 81.2\%  & 72.9\% \\ \hline
\end{tabular}
\end{center}
\begin{flushleft}
$ \ \ \ \ \ \ \ \ \ \ \ \ \ $ \circled{1} Share of the number of submitted CBs (\%).

$ \ \ \ \ \ \ \ \ \ \ \ \ \ $ \circled{2} Share of the number of cleared CBs (\%).

$ \ \ \ \ \ \ \ \ \ \ \ \ \ $ \circled{3} Share of the total submitted quantity in MWh (\%).

$ \ \ \ \ \ \ \ \ \ \ \ \ \ $ \circled{4} Share of the total cleared quantity in MWh (\%).
\end{flushleft}
\end{table}
\color{black}

The total number of \emph{market participants} that ever submitted a CB during the three-year period of this study was 101, with a monthly average of 52 market participants. The total profit that was earned by all the market participants in the CB market during this period was \$61 million.
Out of the 101 convergence bidders, 74 of them made money, i.e., had a \emph{net positive profit}.
Fig. \ref{fig:mon_energy} shows the total monthly amount of cleared energy at each year for the convergence bidders, and Fig. \ref{fig:mon_prof} shows the monthly net profit that all the convergence bidders earned during this period of study.
\color{black}
Note that, notation \$M means \$1,000,000. 
\color{black}
Interestingly, there were months that the market participants had an overall loss, i.e., \emph{net negative profit} as the outcome of their convergence bidding. 
Another interesting observation is that even though the net profit fluctuated significantly across different months, the amount of cleared CB was about the same in each month.

\color{black}
As it is already widely discussed in the literature, there is a direct relationship between the CB market participants' ability to earn profit, and the advantages that convergence bidding can provide to the society from the system viewpoint. In particular, as discussed in the ISO reports, such as in \cite{Ref_CAISO_CB_Document,2015Virtual0Pjm}, if the CB market participants make profit, then their CBs also help closing the gap between D-LMPs and R-LMPs.
Closing such gap results in several benefits to the system, such as \cite{Ref_CAISO_CB_Document,2015Virtual0Pjm}: 
1) Lowering the costs due to more efficient day-ahead commitment,
2) Improving the grid operations and reliability,
3) Market power mitigation,
4) Increasing the market liquidity,
5) Promoting the competition between market participants.
The above mentioned fact in the ISO reports, that a profitable CB helps with achieving price convergence and its advantages, is also proved mathematically in \cite{samani2020cb}.

\subsection{Identifying the Most Present Convergence Bidders} \label{sec:gnral:prsnt}
In this work, although we analyze \emph{all} the submitted CBs in the California ISO market, we scrutinize only the ``most present" convergence bidders for the purpose of extracting their convergence bidding strategies. 
The most present CB market participants can be defined based on different metrics:
1) their high share in the market in terms of the number of submitted CBs;
2) their high share in the market in terms of the number of cleared CBs;
3) their high share in the market in terms of the total amount of quantity of the submitted CBs in MWh; 
4) their high share in the market in terms of the total amount of quantity of the cleared CBs in MWh.

The process of selecting the most present market participants is summarized in Table I. For each metric, we calculated the share (in percentage) of all the CB market participants according to that particular metric. 
We then sorted the list of market participants based on each metric, and accordingly selected the market participants with the 10 highest shares in the market according to each metric. The 10 selected market participants for each metric are marked in Table I by gray shaded areas. There are exactly 10 market participants with gray shaded areas in each column.
Next, we combined the four lists from  the four metrics. Due to the overlaps between the lists for the four metrics, this analysis results in identifying a total of 20 market participants as the ones that are ``most present'' in the CB market.
At this stage, we assigned \emph{Alias IDs} to the selected market participants, as denoted by 1 to 20.
As it is mentioned before, the total number of CB market participants that ever submitted a CB during the period of this study is 101. The reason for choosing the 10 highest shares of each list of metrics is to select all market participants that \emph{one way or another} have some considerable presence in the market, then we scrutinize the selected market participants. 
Note that, each of the 20 selected market participants has a considerable presence in the market based on \emph{at least} one of the four metrics.
As we can see in the last row in Table I, the identified 20 most present market participants in the above process accounted for 72\% to 84\% of the entire convergence bidding market, based on any of the four metrics that one can consider to define the share of the market participants.
\color{black}

\color{black}
Table \ref{tab2} shows some basic information for each Alias ID that we previously identified in Table I. 
\color{black}
We can make several preliminary observations, as we explain next.

Some of the identified convergence bidders placed CBs in almost all the locations that ever received CBs,
such as Alias ID 4 that placed CBs in 95\% of locations that hosted at least one CB at any time during the three-year period of this study.
Some other convergence bidders placed CBs in only a few locations, such as
Alias ID 20 that placed CBs in less than 1\% of the locations.
Most of these 20 convergence bidders with considerable presence submitted both supply and demand bids, but some of them, such as Alias ID 17 and Alias ID 20, submitted supply CBs more than demand CBs, or vice versa.



\color{black}

Based on the average value for the number of steps for the cleared CBs, some market participants always submitted single step bids, while some others used multiple steps in their CBs.
Finally, the average quantity of the cleared CBs varies from about 2 MWh to 156 MWh which shows a different amount of investment and available credit between market participants.

\begin {table}[t]
\color{black}
\centering
\caption {Convergence Bidding Characteristics in the Cleared \\ CBs for the Market Participants with Considerable Presence.}
 \label{tab2}
\begin{center}
   \begin{tabular}{| c | c | c | c | c |}
   \hline

Alias ID & \circled{1} & \circled{2}  & \circled{3} & \circled{4} \\ \hline
1 & 28.21 & 68.11 & 2.86  & 28.79 \\ \hline
2 & 58.53 & 79.08 & 1.11  & 8.41  \\ \hline
3 & 70.95 & 69.33 & 2.05  & 4.92  \\ \hline
4 & 95.37 & 50.53 & 1 & 7.97  \\ \hline
5 & 21.47 & 63.18 & 2.9   & 76.51 \\ \hline
6 & 48.63 & 57.01 & 1.92  & 49.27 \\ \hline
7 & 62.11 & 61.5  & 2.45  & 8.07  \\ \hline
8 & 19.16 & 49.12 & 1.64  & 19.73 \\ \hline
9 & 18.11 & 55.08 & 2.52  & 49.66 \\ \hline
10 & 32.63 & 79.03 & 1.04  & 4.97  \\ \hline
11 & 68.63 & 45.74 & 1 & 1.92  \\ \hline
12 & 20 & 58.02 & 1.03  & 7.84  \\ \hline
13 & 3.58  & 53.81 & 1.37  & 49.15 \\ \hline
14 & 24.84 & 60.19 & 1.13  & 2.6   \\ \hline
15 & 45.26 & 39.25 & 1 & 2.64  \\ \hline
16 & 22.74 & 77.88 & 4.36  & 21.42 \\ \hline
17 & 12 & 87.04 & 3.33  & 53.93 \\ \hline
18 & 33.68 & 53.63 & 1 & 36.23 \\ \hline
19 & 69.89 & 64.09 & 5.29  & 37.37 \\ \hline
20 & 0.21  & 100   & 1 & 156.72 \\ \hline

\end{tabular}
\end{center}
\begin{flushleft}
$ \ \ \ \ \ \ \ \ \ \ \ \ \ $ \circled{1} Share of nodal locations in cleared CBs (\%).

$ \ \ \ \ \ \ \ \ \ \ \ \ \ $ \circled{2} Share of supply bids in cleared CBs (\%).

$ \ \ \ \ \ \ \ \ \ \ \ \ \ $ \circled{3} Average number of steps in cleared CBs.

$ \ \ \ \ \ \ \ \ \ \ \ \ \ $ \circled{4} Average quantity in MWh in cleared CBs.
\end{flushleft}
\end{table}
\color{black}

\section{Data-Driven Reverse Engineering of the Convergence Bidding Strategies} \label{sec:classes}

In this section, first, we will extract different quantitative \emph{features} to characterize the convergence bidding strategies of the market participants based on the raw market data that we introduced in Section II.
After that, we will use the extracted features to \emph{cluster} the submitted CBs into three clusters of strategies.
Finally, the performance of the clusters of strategies will be compared. By going through these steps, the convergence bidding strategies of the real-world market participants in the California ISO market will be \emph{reverse engineered}.
\color{black}
The benefit of this analysis is two-fold.
First, an in-depth understanding of the CB strategies that are currently adopted by the real-world market participants is in its own right interesting from the view point of research and also to provide insights to ISOs.
Second, by unmasking and reverse engineering the existing real-world CB strategies in the California ISO market, an enhanced and more profitable CB strategy is achieved, as we will see in Section IV. 
\color{black}

\subsection{Features for Cluster Identification}
\label{sec:features}

Recall from Section II.A that each submitted CB has three types of information: quantities, prices, and whether it is a demand CB or a supply CB. The pair of quantity and price can be submitted in one step or multiple steps.
Accordingly, we introduce four different \emph{features} for each submitted CB: 
1) The price distance, i.e., the difference between the price bid in the submitted CB at a node and the average hourly D-LMP\footnote{\color{black}The average D-LMP is a fixed number for each hour and each node, as it is the mean value of the historical prices over a period of three years. \color{black}
}
at that node;
2) The correlation between the type of the submitted CB (demand or supply) at a node and the historical CBs at that node;
3) The number of steps in the submitted CB;
4) The type of node where the CB is submitted, i.e., whether the
node is a regular APnode, \color{black} i.e., it is \emph{not} a Hub or a DLAP, \color{black} or it is one of the major aggregated
nodes in the California ISO market, \color{black} i.e., it \emph{is} a Hub or a DLAP\color{black}.

\color{black}
In this work, we seek to consider the key determinative features for the bidding strategy of the CB market participants.
In our assessment, the selected features should have three main characteristics as follows. 
First, the selected feature should be built only based on the data that each market participant has access to by its own. For example, a feature that needs to include other market participants' bidding information is not considered in our features for the purpose of clustering.
\color{black}
This is because each convergence bidder does not have access to the bidding data of other market participants.
\color{black}
Second, the selected feature should not involve or depend on the information that is private to the market participant.
\color{black}
It is not really a choice to not consider such private information; it is rather the nature of a study like ours that is based on analyzing real-world electricity market data.
\color{black}
For example, each CB market participant has a ``credit'' with California ISO, which determines the maximum quantity of bids that the market participant can submit to the market. Such ``credit'' is not public data. Thus, 
the quantity of the submitted CB in MWh is not considered as a feature, because it is not clear whether the quantity of the bid is simply set based on the market participant's ``credit'' or it is a factor that is strategically selected by the market participant.
Third, the selected feature should be built only based on the data that each market participant has access to \emph{at the time of submitting} its bid to the California ISO market. For example, the same day LMPs are \emph{not} known to the market participants at the time of submitting their bids, but the historical average of the LMPs for each location and each hour \emph{is} known to them.

\begin{figure}[t]
 \centering
{\scalebox{0.42}{\includegraphics*{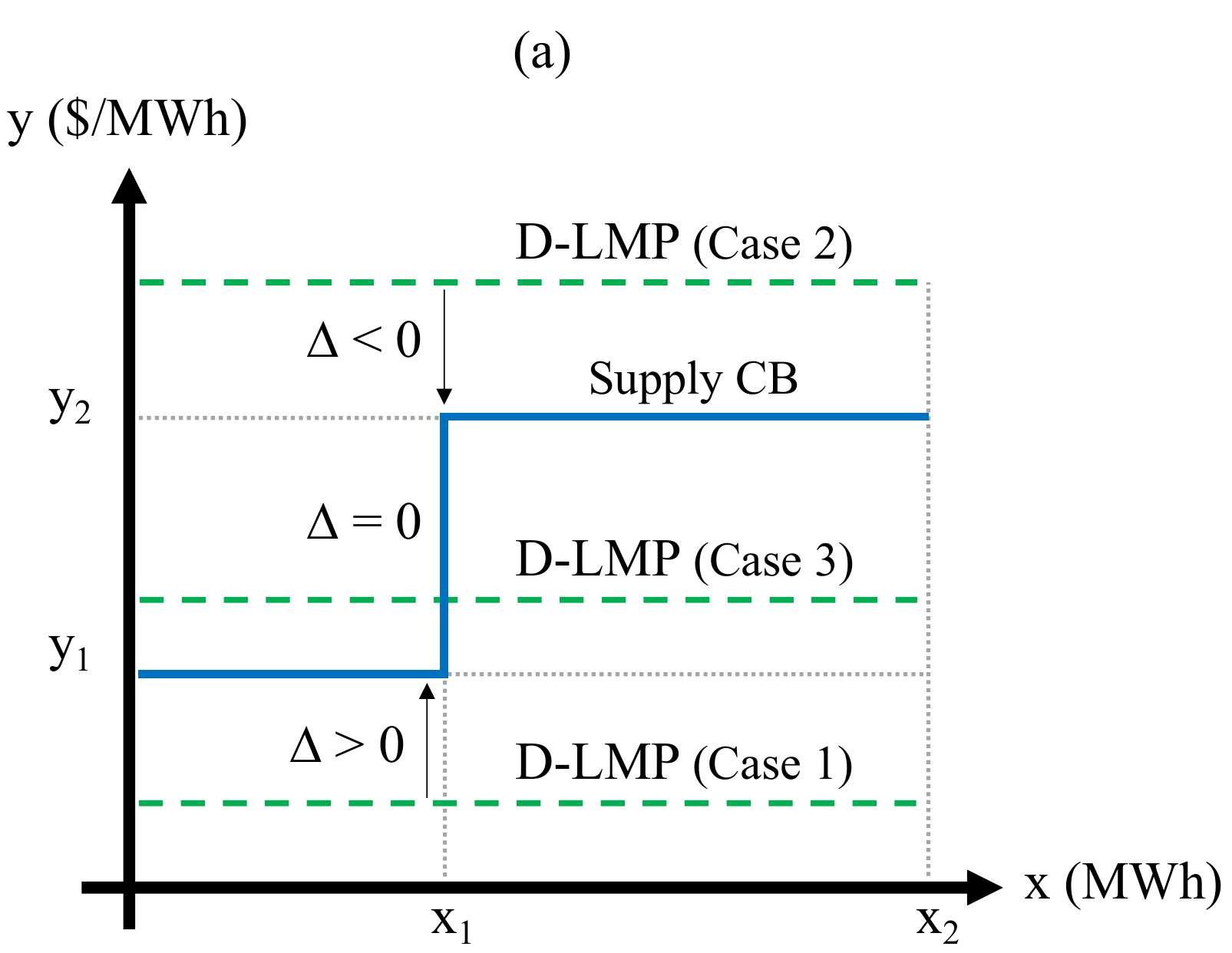}}}
{\scalebox{0.42}{\includegraphics*{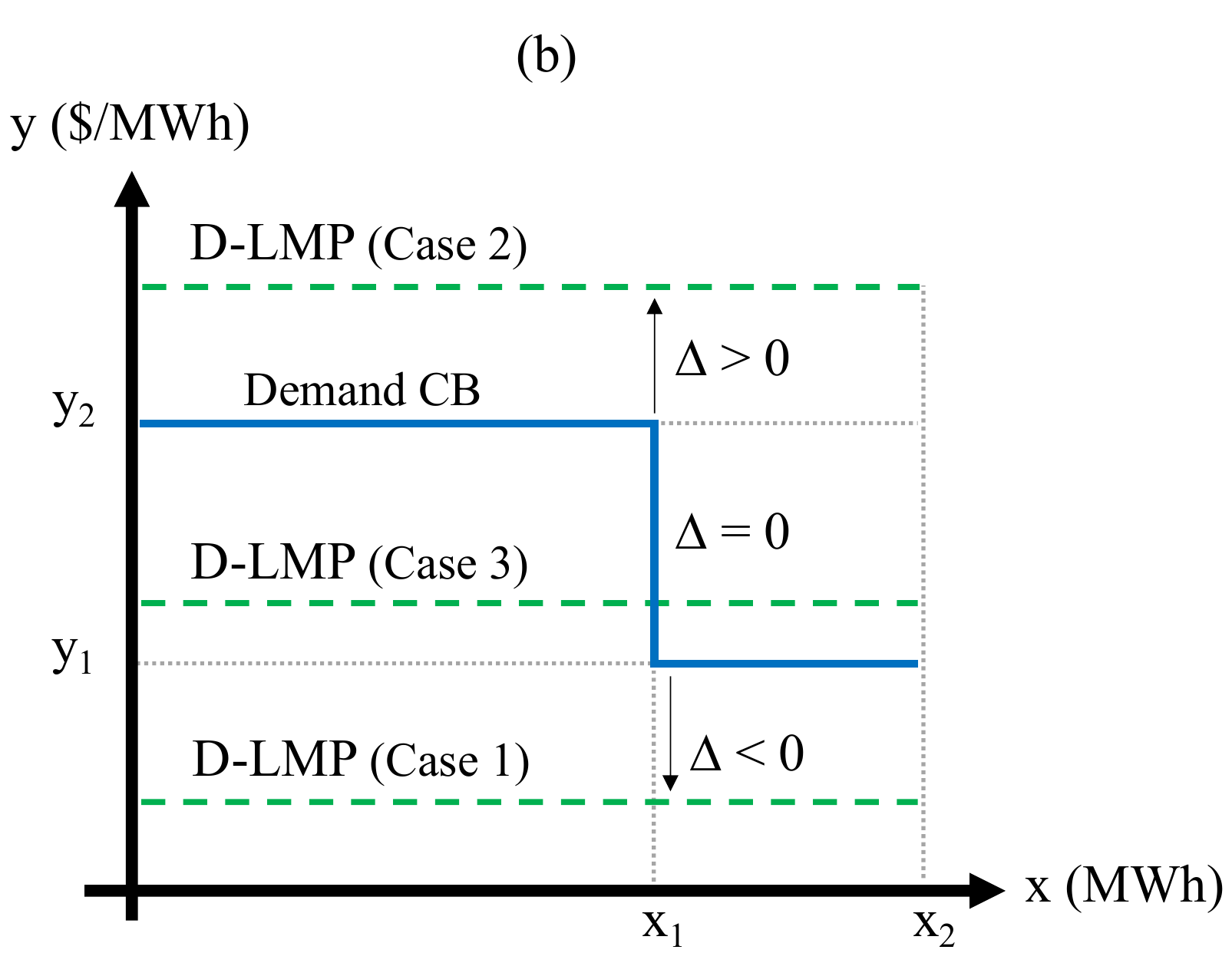}}}
\color{black}
\caption{Demonstration of the first feature, i.e., $\Delta$, which is the price distance of bid from average hourly D-LMP: a) in a supply bid; b) in a demand bid.}
\color{black}
\label{fig:dist_def}
\end{figure}

\color{black}
Fig. 4 shows the definition of the first feature for a multi-step supply CB and a multi-step demand CB. This feature is denoted by $\Delta$.
\color{black}
In this figure, the green horizontal dashed line is the average hourly D-LMPs for the hour corresponding to a given CB. Three green lines show three possible different cases that will be explained in follow. 
\color{black}
First, consider a supply CB \color{black} in sub-figure (a)\color{black}. Three cases can happen:
1) All the submitted price values in a given CB are higher than the average D-LMP. In other words, the entire piecewise linear function for the submitted CB is above the average D-LMP.
In this case, \color{black} $\Delta > 0$ and it \color{black} is defined to be equal to the minimum price value in the submitted CB minus the average D-LMP. 
2) All the submitted price values in a given CB are lower than the average D-LMP. In other words, the entire piecewise linear function for the submitted CB is below the average D-LMP. 
In this case, \color{black} $\Delta < 0$ and it \color{black} is defined to be equal to the maximum price value in the submitted CB minus the average D-LMP. 
3) The average D-LMP is somewhere between the minimum and the maximum price values in the submitted CB. In other words, the average D-LMP has an intersection with the piecewise linear function for the submitted CB.
In this case, \color{black} $\Delta = 0$ and it \color{black} is defined to be zero. 
Next, consider a demand CB \color{black} in sub-figure (b)\color{black}. Again, three  cases can happen, which can be defined similarly. The only difference is that, when it comes to a demand CB, in the first case, $\Delta$ is equal to the average D-LMP minus the minimum price value in the submitted CB (not the other way around); and in the second case, $\Delta$ is  equal to the average D-LMP minus the maximum price value in the submitted CB (not the other way around). In other words, the previously defined $\Delta$ should be multiplied by -1.
\color{black}

The second feature indicates whether the same type of CB, i.e. a supply CB or a demand CB, has been consistently used by a market participant at a  nodal location.
This feature is a number between $0$ and $1$. 
\color{black}
This feature indicates whether the type of the submitted CB is similar or dissimilar to the type of the CBs historically submitted by the same market participant at the same location. 
The value of this feature is close to $1$, if the market participant consistently selects the same type of CB at the given node; and the new submitted CB also has the same type.
The value of this feature is close to $0$, if the market participant consistently selects the same type of CB at the given node; but the new submitted CB has a different type.
Finally, the value of this feature is close to $0.5$ if the market participant frequently changes the type of its submitted CBs at the given node; i.e., it submits a mix of both supply CBs and demand CBs.
\color{black}As an example, consider all the previous CBs that are submitted by a market participant at a node. Suppose 60\% of the CBs are supply bids and 40\% of the CBs are demand bids. If the current CB is a supply CB, then the second feature would be 0.6. If the current CB is a demand CB, then the second feature would be 0.4.
\color{black}
In this work, a window of one year of historical data, i.e., the data over the previous year, is considered for calculating this feature.
\color{black}

The third feature is the number of steps in the submitted CB, which is an integer number between one and ten.

The fourth feature is driven by the fact that some market participants submit CBs \emph{only} at the major aggregated nodes, in the California ISO market, 
i.e., at one or more of its three Hubs, namely NP15, SP15, and ZP26, or its three Default Load Aggregated Points (DLAPs), which include San Diego Gas and Electric (SDG\&E), Pacific Gas and Electric (PG\&E), and Southern California Edison (SCE).
Importantly, these major aggregated nodes have \emph{a higher level of predictability} for LMPs, compared to the regular APnodes.
\color{black}
This feature somewhat incorporated two other candidate features, namely the LMP volatility and the LMP forecast accuracy of the node where the CB is submitted. In fact, higher volatility in LMP values directly results in less accuracy in forecasting the LMP values.
This fourth feature is a binary number on whether or not the node is a major aggregated node.
It should be noted that, in the California ISO market, the major aggregated nodes are among the APnodes.
\color{black}
\subsection{Identified Convergence Bidding Clusters}
Based on the introduced features in Section III.A, next, we classify the submitted CBs  by using the Hierarchical Density-Based Spatial Clustering of Applications with Noise (HDBSCAN) method. HDBSCAN is a robust clustering algorithm that can work with little or no parameter tuning \cite{campello2013density}. The only parameter that needs to be tuned in this method is the minimum number of points in each cluster.

\color{black}
The data points in this analysis are the collection of all the submitted CBs over the three-year period of this study, which add up to 6.6 million CBs.
The purpose of our clustering analysis is to gain insights from such a huge amount of data, such that we can identify the main convergence bidding strategies in the California ISO market. 
Accordingly, our analysis is a hybrid of applying data-driven algorithms and manual inspection of the data-driven results.
%
\color{black}
In the latter (i.e., manual) inspection, we examined the data-driven results with respect to the four features and we accordingly identified
%
\color{black}
three clusters that can cover practically all the existing major convergence bidding strategies in the California ISO market during the period of this study.
%
In this process, we combined artificial intelligence with human expertise to translate the bidding data to the most meaningful clusters.
It should be mentioned that, \color{black} the clustering in this paper is done only once and it is done for the entire dataset in the period of this study. Our approach to involve
\color{black}
both machine intelligence and human expertise, is very suitable for the purpose of this study which involves a huge amount of bidding data.
The definition of each cluster and its work function is defined as follows:

\vspace{0.05cm}
\color{black}\textbf{CB Cluster 1 (Price-Forecasting Strategy):}
This strategy is the case where the CB market participant submits price bids that are \emph{close} to the LMP values at the location where the CB is placed; making it evident that the market participant is trying to forecast the market prices at its bidding locations. 
For each hour of the next day, if the forecasted D-LMP is higher than R-LMP, then a supply CB is submitted.
If the forecasted D-LMP is lower than R-LMP, then a demand CB is submitted.
For a supply CB, the price values should be \emph{less} than the forecasted D-LMP but \emph{close} to it in order to avoid entering the market when D-LMP is unexpectedly low. Also, for a demand CB, the price values should be \emph{more} than the forecasted D-LMP but \emph{close} to it in order to avoid entering the market when D-LMP is unexpectedly high.
As a result, in this cluster:
1) $\Delta$ is relatively small.
2) The correlation between the type of the submitted CB and those of the previous CBs of the same market participant at the same location is often {not} close to 1, because the convergence bidder is trying to actively forecast the LMPs. As a result, the types of the CBs are selected according to the forecast results, and they can vary depending on the market conditions.   
3) The number of steps for the submitted CB can be single or multiple, and this is \emph{not} a determinative feature in this cluster of strategies.
4) The CB is mostly submitted in a major aggregated node, with a higher level of locational price predictability, because this strategy requires accurate forecasting of both D-LMP and R-LMP. However, in principle, it is possible that a market participant uses this strategy on regular APnodes, if they \emph{can} achieve accurate LMP forecasts at that node.

\vspace{0.05cm}
\textbf{CB Cluster 2 (Self-Scheduling Strategy):}
This strategy is the case where the CB market participant does not mean to calculate and submit a price bid, i.e., its CB is mainly about its quantity.
It should be clarified that, in principle, all CBs in the California ISO market \emph{must} include at least one price value.
Thus, when a CB market participant follows a self-scheduling strategy, it still needs to include a price value in its CB.
For a demand CB, the price value should be \emph{much higher} than the expected D-LMP for that hour, i.e., the average D-LMP for that hour, such that the submitted CB is always cleared in the market.
For a supply CB, the price value should be \emph{much lower} than the expected D-LMP, such that the submitted CB is always cleared in the market. 
In both cases, i.e., whether the submitted CB is a demand bid or a supply bid, it would result in a large negative $\Delta$.
Importantly, as far as the price-forecasting is concerned, the self-scheduling strategy only needs a \emph{rough forecast} about the \emph{sign} of the difference between D-LMP and R-LMP in order to decide on whether to submit a supply CB or a demand CB.
For each hour of the next day, if the difference between D-LMP and R-LMP is expected to be \emph{positive}, then a supply CB is submitted; and if such difference is expected to be \emph{negative}, then a demand CB is submitted.
As a result, in this cluster:
1) $\Delta$ is relatively large and negative.
2) The correlation between the type of the submitted CB and those of the previous CBs  of the same market participant at the same location is often \emph{not} close to 1, because the convergence bidder may submit different types of CBs based on the expected sign of the difference between D-LMP and R-LMP.
3) The submitted CB is \emph{single} step. This is an important determinative feature in this cluster of strategies, because multi-step strategies cannot match the definition of self-scheduling bids. 
4) The CB may be submitted at regular APnodes \emph{or} at the major aggregated nodes.

\vspace{0.05cm}
\textbf{CB Cluster 3 (Opportunistic Strategy):}
This strategy is the case where the CB market participant does not want to get involved in the difficulties of doing an accurate price forecast, yet it does not want to be as passive as in the self-scheduling strategy (as far as the selection of its price bid is concerned). Hence, the market participant takes a third option, which is somewhat opportunistic. 
In this strategy, the CB market participant always submits either a supply CB that has a price bid that is considerably higher than the
\color{black}
D-LMPs\color{black},
or a demand CB that has a price bid that is considerably lower than the \color{black} D-LMPs\color{black}.
%
In this regard, the CB market participant waits for a \emph{spike} in D-LMP to enter the market. 
As a result, the submitted bids are \emph{not} cleared most of the time. They are cleared only occasionally, when there is a \color{black}potential \color{black} opportunity to make a considerable profit.
Interestingly, this is a completely new CB strategy and it does not match any of the strategic convergence bidding methods that currently exist in the literature.
Thus, we will provide a detailed explanation about the application of this strategy in Section IV.
In this cluster:
1) $\Delta$ is relatively large and positive.
2) The correlation between the type of the submitted CB and those of the previous CBs \emph{is} often close to 1. 
3) The number of steps for the submitted CB can be single or multiple, and it is \emph{not} a determinative feature in this cluster.
4) The CB is almost always submitted at regular APnodes, but \emph{not} at the major aggregated nodes. This is an important determinative feature in this cluster, because convergence bidders should find those nodes that have a potential for experiencing price spikes. Major aggregated nodes with high levels of predictability do not carry this characteristic.

\color{black}
The above three identified clusters and their introduced characteristics address Research Question 2. 

The above analysis can also be used to address Research Question 3. Importantly, while \color{black} CB Cluster 2 is \emph{less} common among market participants, yet many of the existing papers in the literature \emph{are} in fact focused on this strategy; such as some of the papers that we cited in Section I.C.
Furthermore, CB Cluster 3 is a common strategy in the California ISO market, yet it does \emph{not} match any of the strategic convergence bidding methods that currently exist in the literature. In fact, the strategy in CB Cluster 3 is currently used by several market participants, including Alias ID 1, which is the most active market participant in the California ISO market, in terms of the number of submitted CBs. 

\begin{figure}[t]
 \centering
{\scalebox{0.45}{\includegraphics*{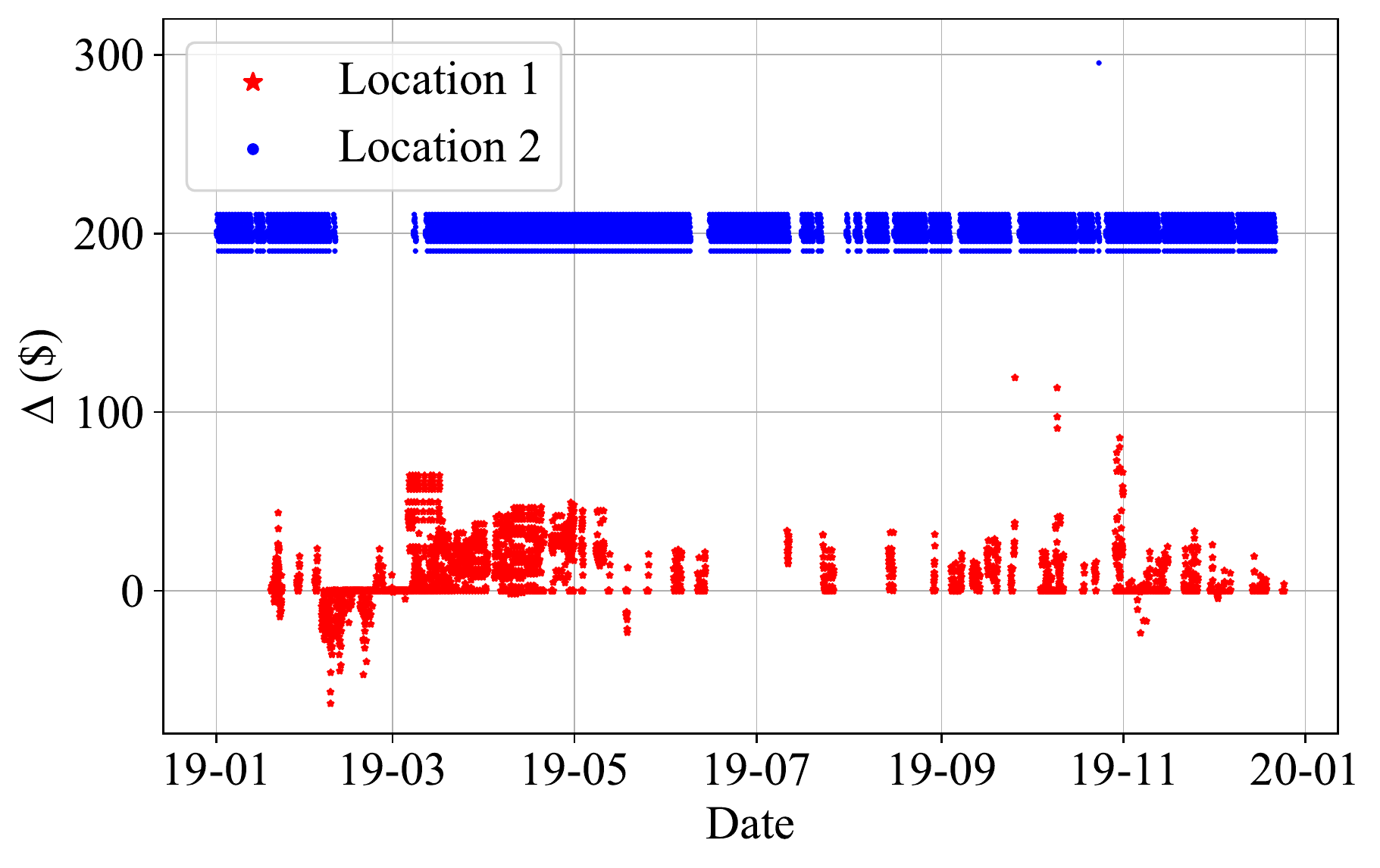}}}
\caption{The first feature ($\Delta$) for Alias ID 5 at two different locations during 2019. It is evident that this CB market participant uses two different bidding strategies as these two different locations.}
\label{fig:214147_dist}
\end{figure}

Another note to highlight is that the \emph{same} market participant may have \emph{different} strategies at \emph{different} nodes. For example, consider the hourly value of the first feature ($\Delta$) in Fig. \ref{fig:214147_dist} for the submitted CBs by Alias ID 5 in two \emph{different} locations. 
As we can see, there is a clear distinction between the strategies that Alias ID 5 chose at these two different locations.

\begin{figure}[t]
 \centering
{\scalebox{0.45}{\includegraphics*{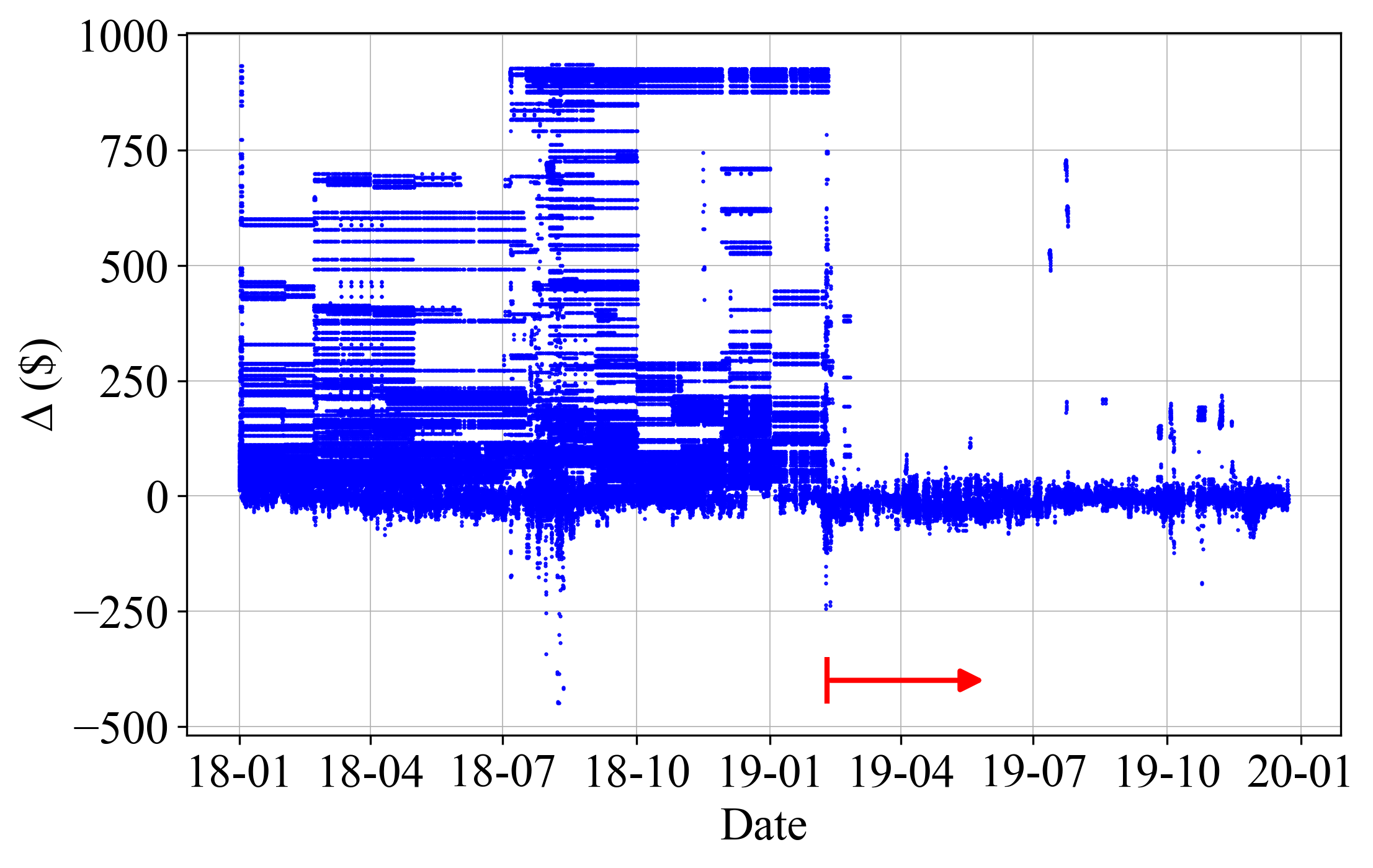}}}
\color{black}
\caption{The first feature ($\Delta$) at each hour for Alias ID 6 over a period of two years. It is evident that this CB market participant changed its bidding strategy in February 2019.}
\color{black}
\label{fig:chng_str}
\end{figure}

\color{black}
An interesting observation is that, some of the CB market participants have clearly \emph{changed} their strategy during the period of this study. For example, Fig. \ref{fig:chng_str} shows the first introduced feature, i.e., the distance of the submitted price bids from the average D-LMPs, at each hour for Alias ID 6. We see that this market participant clearly changed its bidding strategy around February 2019. While its bidding strategy in 2018 mostly matches the Opportunistic Strategy, its bidding strategy in 2019 mostly matches the Price-Forecasting Strategy.

\color{black}As a side note, no other market participant changed her convergence bidding strategy around the date that Alias ID 6 changed her strategy. Therefore, while we cannot speculate on the reason for Alias ID 6 to change her strategy, it is more likely that the change was due to Alias ID 6’s own internal factors than a  change in the system.
\color{black}

\subsection{Performance Comparison among Identified Strategies} \label{sec:performance}
To complete the reverse engineering task, next, we evaluate the performance of each CB cluster to understand the advantages and the disadvantages of different CB strategies.
\color{black}
Two metrics are used to assess and compare the performance of different CB clusters. 
The first metric is the \emph{cleared-to-submitted-ratio} (CSR), which is the percentage of the submitted CBs that are cleared in the market for each market participant:
\begin{equation} \text{CSR} = \frac{\text{Number of Cleared CBs}}{\text{Number of Submitted CBs}} \times 100.
\end{equation}

The second metric is the \emph{loss-to-profit-ratio} (LPR), which can help capture the level of loss compared to the level of profit. 
This metric is defined as follows:
%
\begin{equation} \label{define_LPR}
\text{LPR} = \frac{\text{Total Loss}}{\text{Total Profit}} \times 100.
\end{equation}

A lower CSR indicates that only a small portion of the submitted CBs for a given market participant at a given node is cleared.
A lower LPR indicates that the cleared CBs of a given market participant at a given node resulted in more profits than losses. 
Together, CSR and LPR draw a clear picture about the portion of the CBs that are cleared and the circumstances in terms of loss versus profit for the cleared CBs.
\color{black} 
Note that, we do \emph{not} consider the total net profit as a comparison factor, because it depends on each market participant's credit in the California ISO market, which limits the quantity of their submitted CBs.

\begin {table}[t]
\color{black}
\centering
\caption {Percentage of Submitted CBs that Are Cleared in the \\ Market for Each Alias ID in Each Year.}
\label{tab3}
\begin{center}
  \begin{tabular}{| c | c | c | c |}
  \hline
Alias ID & 2017 (\%)  & 2018 (\%)   & 2019 (\%)  \\ \hline
1    & 3.58 & 3.59 & 2.45 \\ \hline
2    & 66.12 & 65.72 & 67.36 \\ \hline
3    & 68.43 & 68.86 & 48.32 \\ \hline
4    & -    & 98.68 & 94.38 \\ \hline
5    & 33.46 & 12.54 & 5.25 \\ \hline
6    & 7.29 & 14.41 & 44.49 \\ \hline
7    & 5.52 & 13.48 & 11.51 \\ \hline
8    & 100  & 4.54 & 14.23 \\ \hline
9    & 88.36 & 94.15 & 76.77 \\ \hline
10   & 56.99 & 60.57 & 40.17 \\ \hline
11   & -    & 99.84 & 99.98 \\ \hline
12   & -    & 61.14 & 54.99 \\ \hline
13   & 61.14 & 51.78 & 51.12 \\ \hline
14   & 98.19 & 93.11 & 95.54 \\ \hline
15   & 96.45 & 99.9 & 99.79 \\ \hline
16   & 68.81 & 81   & 75.64 \\ \hline
17   & 86.43 & 88.15 & 49.98 \\ \hline
18   & 82   & 59.73 & 33.62 \\ \hline
19   & 78.45 & 83.99 & 84.15 \\ \hline
20   & 98.79 & 98.13 & -    \\ \hline
\end{tabular}
\end{center}
\end{table}
\color{black}
The CSR is listed in Table \ref{tab3} for all the identified Alias IDs. 
\color{black}
For each CB cluster, the performance of one representative market participant is considered for benchmarking.

\begin{figure}[t]
\color{black}
 \centering
{\scalebox{0.45}{\includegraphics*{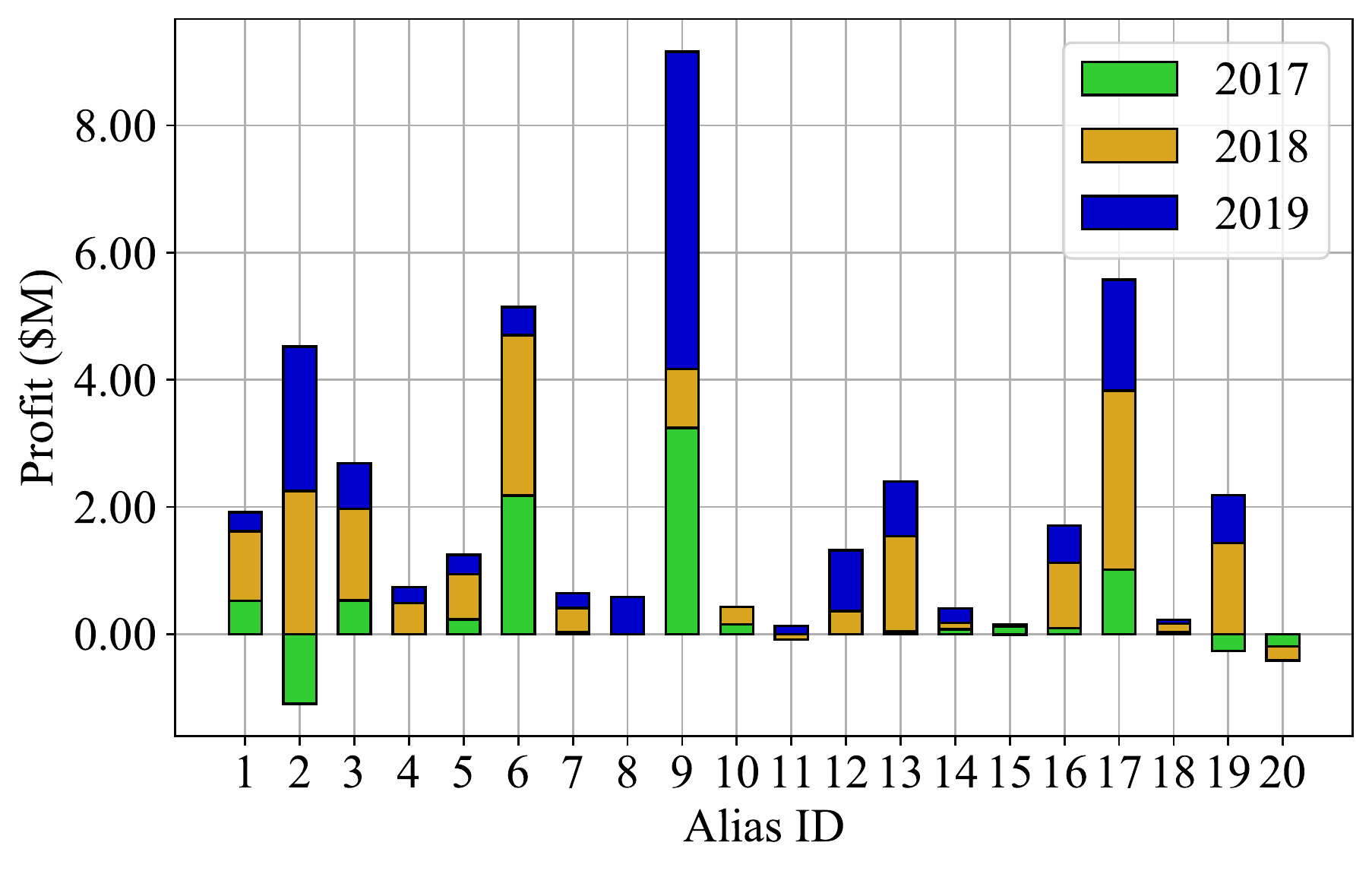}}}
\caption{Total yearly amount of earned/lost profit by each Alias ID.}
\label{fig:totl_prft_id}
\end{figure}
\color{black}

\begin{figure}[t]
 \centering
{\scalebox{0.45}{\includegraphics*{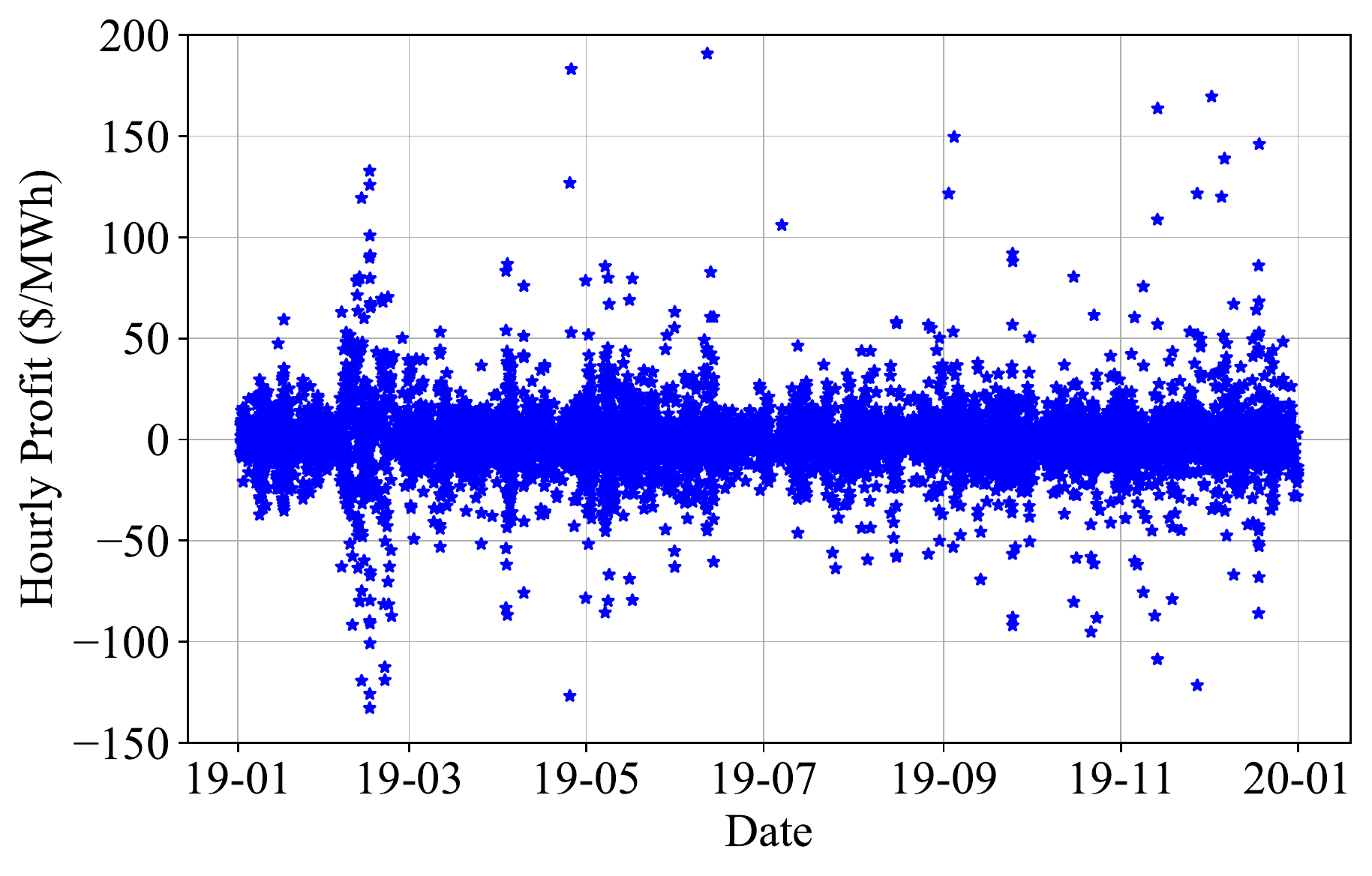}}}
\caption{Hourly profit for Alias ID 9 using the Price-Forecasting Strategy at a DLAP.}
\label{fig:SDGE_990504}
\end{figure}

\color{black}
Alias ID 9, which is the \emph{most lucrative} CB market participant in the California ISO market (Fig. \ref{fig:totl_prft_id}), mostly used the Price-Forecasting Strategy on the major APnodes.
Fig. \ref{fig:SDGE_990504} shows the hourly per unit profit (\$/MWh) for Alias ID 9 in one of the three DLAPs using the Price-Forecasting Strategy.
As we can see, there are many hours with a loss for the submitted CBs by this market participant.
For these CBs, CSR and LPR are 86.43\% and 73.62\%, respectively. These values show that most of the submitted CBs by Alias ID 9 are cleared, and some of the cleared ones resulted in a positive profit.  
\color{black}

\begin{figure}[t]
 \centering
{\scalebox{0.45}{\includegraphics*{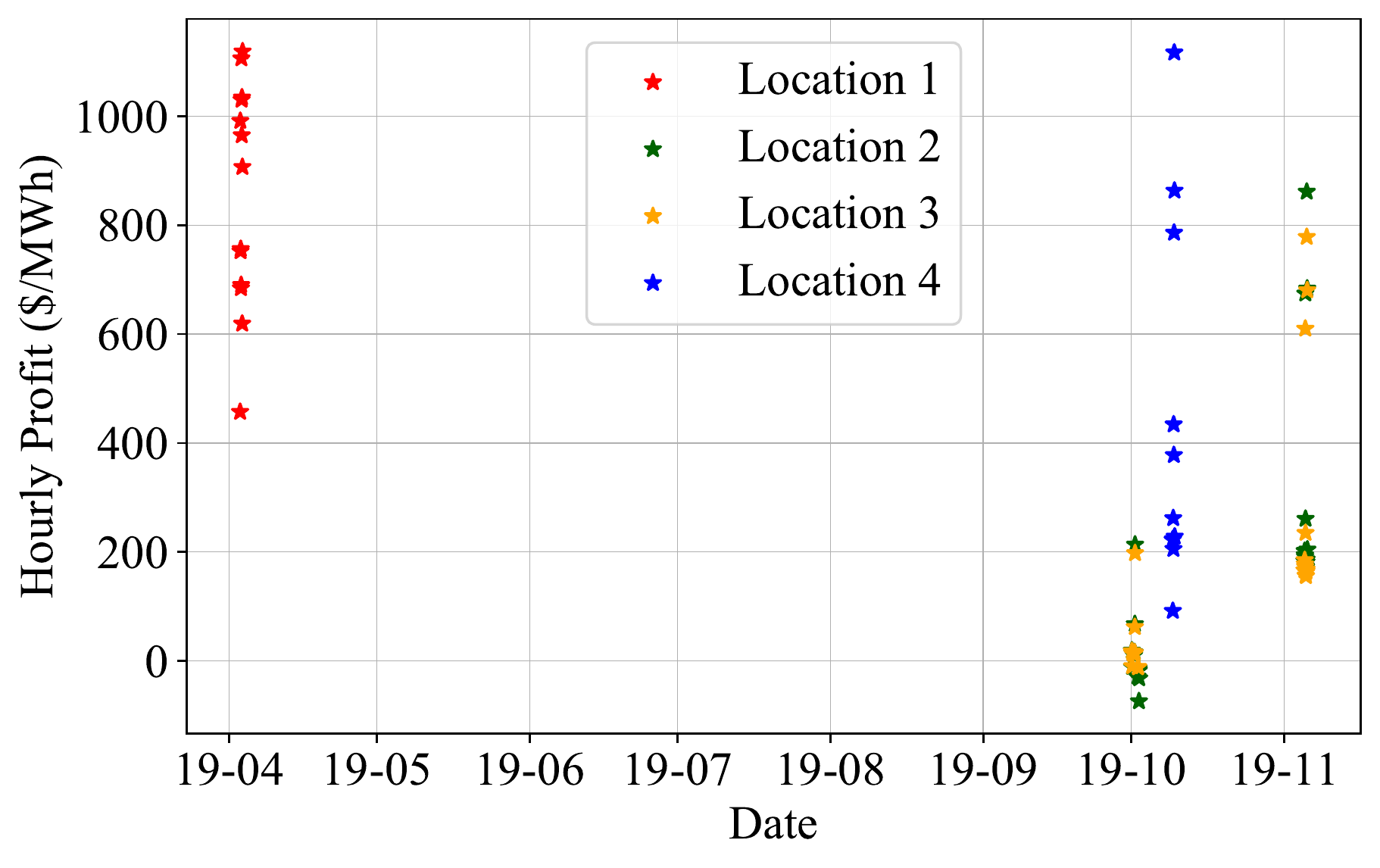}}}
\caption{Hourly profit for Alias ID 1 using the Opportunistic Strategy at four different locations in the market.}
\label{fig:774727}
\end{figure}

\color{black}

\begin{figure}[t]
\color{black}
 \centering
{\scalebox{0.45}{\includegraphics*{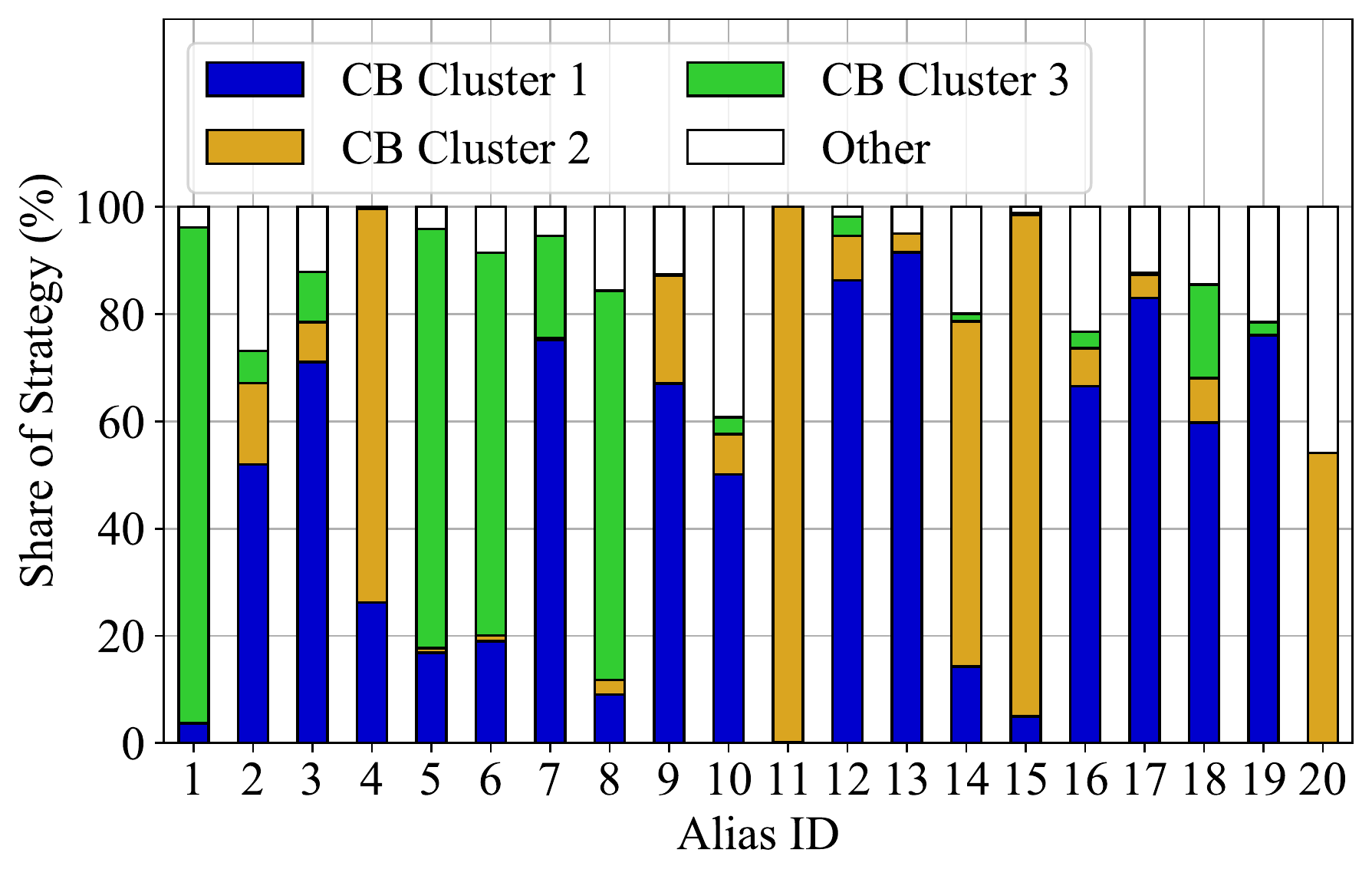}}}
\caption{The share of each identified strategy for each market participant.}
\label{fig:share_strtgy}
\end{figure}
\color{black}

Alias ID 1, which is the most \emph{active} market participant in terms of the number of submitted CBs, used the Opportunistic Strategy in \emph{most} of the nodes.
Fig. \ref{fig:774727} shows the hourly per unit profit for Alias ID 1 at four different locations using the Opportunistic Strategy. It should be mentioned that only the non-zero profits are shown in Fig. \ref{fig:774727}.
As we can see, there are only a few days that \emph{any} of the submitted CBs is cleared. But Alias ID 1 had excellent profit on those few days.
\color{black}
For these CBs, CSR and LPR are 0.24\% and 3.04\%, respectively. These values show that only a few submitted CBs are cleared, but most of the cleared ones resulted in a positive profit.  
\color{black}

Alias ID 11 always used the Self-Scheduling Strategy as its strategy. As mentioned in Section III.B, this strategy is \emph{less} common among the market participants.
Fig. \ref{fig:totl_prft_id} shows that Alias ID 11 did \emph{not} gain a high profit during its presence in the market despite participating in more than 300 nodes.
From Table \ref{tab3}, almost 100\% of Alias ID 11's CBs are cleared in the market.
\color{black}
For this market participant, CSR is 90.90\%.

\color{black}
The share of each implemented CB strategy for each CB market participant is calculated. The results for each of the 20 most present CB market participants are shown in  Fig. 10. This figure shows the share of each strategy for each Alias ID. We can see that each of the three identified convergence bidding strategies has been used in the market.  

In total, out of the 6.6 million submitted CBs that were analyzed during the period of this study for all the 101 CB market participants in the California ISO market, here is the share of each convergence bidding strategy: 35.95\% of all the submitted CBs belong to CB Cluster 1 (Price-Forecasting Strategy), 15.58\% of all the submitted CBs belong to CB Cluster 2 (Self-Scheduling Strategy), 35.33\%  belong to CB Cluster 3 (Opportunistic Strategy), and 13.14\%  of all the submitted CBs belong to Other (Unidentified) strategies.

\color{black}
As it is evident from the above numbers, \emph{three} is the exact number of the clusters of strategies that take \emph{significant shares} of the real-world convergence bids in the period of this study. On one hand, a smaller number of clusters would inevitably ignore at least one of the three significant real-world strategies.
On the other hand, a larger number would inevitably add a very insignificant strategy which can distract the focus from the three significant strategies and their characteristics and implications.

Collectively, the analyses in Sections III-B and III-C address Research Question 4 that we had raised in Section I.A.
\color{black}

\section{Designing a Comprehensive Convergence Bidding Strategy Based on the Reverse Engineering Results} \label{sec:optiz}

In this section, we \color{black} seek to address Research Question 5. In this regard, we \color{black}propose a new comprehensive composite convergence bidding strategy based on the results in Section III.
The key question is: \emph{now that we have learned the strategic behaviors of various real-world CB market participants through reverse engineering, can we go one step further and create a new CB strategy that can learn from the advantages and disadvantages of the existing strategies to significantly outperform them?}
The answer is \emph{yes}. In this section, we discuss how such a strategy can be developed.

The proposed composite convergence bidding strategy is developed in three steps. \textbf{\emph{First}}, we focus on the Opportunistic Strategy, i.e., the CB strategy that is completely new and has never been discussed in the literature.
In this step, we propose an optimization-based algorithm to maximize the net profit of the market participant by \emph{capturing the spikes} in D-LMPs with an optimal price bid at each node.
\emph{\textbf{Second}}, we introduce an algorithm to \emph{label} each node based on the solution of the optimization problem in the first step. This labeling is necessary to find out what kind of CB (if any) is more profitable at each node. Each node can be labeled as Demand CB Node, or Supply CB Node, or Neither, or Both.
\textbf{\emph{Third}}, by using the results from the first two steps, and combining them with the results from Section III about CB Cluster 1 and CB Cluster 2, we propose the new composite CB strategy. 

\subsection{Step 1: Net Profit Maximization by Capturing Price Spikes} \label{sec:step1}

\subsubsection{Basic Idea}
We define \emph{price spikes} as the cases where D-LMPs demonstrate abnormalities by being much higher or much lower than the average D-LMP at the same location and the same hour.
By using the historical data at each node, we formulate an optimization problem to find the \emph{optimal price bid} that could maximize the net profit with \emph{minimum loss} for a CB with a unit quantity. 
If there is a feasible solution for the formulated optimization problem, then the optimal price bid will be used for each hour of the next day for that given node. Price spikes can be both negative or positive. 
A demand CB is required to take advantage of a negative spike in D-LMPs, i.e., low prices. Similarly, a supply CB is required to take advantage of a positive spike in D-LMPs, i.e., high prices. 

\subsubsection{Optimization Problem Formulations}
The following optimization problem is designed to capture the negative spikes with a demand CB at each node in order to maximize the total net profit with minimum loss.
At each time interval, if the net profit ($\eta_t$) is positive, then it is considered as a profit ($P_t$); and if it is negative, then it is considered as a loss ($L_t$).
\color{black}
Note that, this optimization problem makes use of only the historical data that are already known to the market participant.
\color{black}
Thus, it does \emph{not} require dealing with the difficulties associated with price forecasting. Hence, this analysis is inherently not sensitive to the accuracy of price forecasting. 
\color{black}
Here, $T$ is the set of historical time intervals that a market participant considers in analyzing the historical price spikes.
\color{black}
\begin{align}
& {\text{maximize}} && Obj : \sum\limits_{t = 1}^T \eta_t\\
& \text{subject to} && \eta_t = (\pi_t - \lambda_t) \; | \; \lambda_t \leq x_t, &\forall t \in T \\
&  && x_t = {\lambda}_t^{*} - m &\forall t \in T \\ 
&  && P_t = \eta_t \; | \; \eta_t \geq 0, &\forall t \in T \\ 
&  && L_t = \eta_t \; | \; \eta_t \leq 0, &\forall t \in T \\
&  && -\sum\limits_{t = 1}^T L_t \leq \epsilon \times \sum\limits_{t = 1}^T P_t \\
&  &&  m^{min} \leq m \leq m^{max}&&
\end{align} \label{eq:noncnvx}
The objective function (3) is the total net profit over the past $T$ time intervals. The amount of net profit for each time interval using a demand CB with a unit quantity is calculated in (4). Here, $\pi$ and $\lambda$ denote R-LMP and D-LMP, respectively; and $x$ is the price bid.
The \emph{condition} in this constraint, which is denoted by a vertical line, indicates that the submitted price bid ($x$) is cleared \emph{only if} it is higher than D-LMP.
Eq. (5) shows that the price bid for each hour is equal to the average of D-LMP for that hour minus $m$.
In this optimization, $m$ is the main decision variable which is the \emph{distance} from the average hourly D-LMPs that is captured as a \emph{price spike criteria} by an optimal price bid. 
Eqs. (6) and (7) divide the net profit to \emph{profit} and \emph{loss} in each time interval.
Eq. (8) is a bound constraint that is used to guarantee that the total amount of loss is less than a small percentage ($\epsilon$) of the total amount of profit.
As we will discuss in Step 2 of the proposed method, only those nodes that have a \emph{feasible solution} for this optimization problem with an optimal objective value of greater than a \emph{threshold} are used in our proposed bidding strategy.

As mentioned before, the optimization problem in (3)-(9) is for capturing the \emph{negative} spikes in D-LMPs with demand CBs. The same optimization problem can be used to capture the \emph{positive} spikes with supply CBs. We just need to replace the definition of net profit in (4)-(5) with the following:
\begin{align}
&  && \eta_t = (\lambda_t - \pi_t) \; | \; \lambda_t \geq x_t, &\forall t \in T \ \! \\
&  && x_t = {\lambda}_t^{*} + m &\forall t \in T &&
\end{align} \label{eq:spplycb}
\vspace{-.4cm}
\subsubsection{Solving the Formulated Problems}
The introduced optimization problems are \emph{non-convex} and may not be solved efficiently and quickly in their current forms.
Importantly, it is necessary to have a computationally tractable formulation as these optimization problems must be solved \emph{each} day for \emph{all} the nodes in the market.
In two steps, we convert the optimization problem (3)-(9) into a Mixed-Integer Linear Program (MILP).
First, we introduce a binary variable ($b_{t}^1$) and utilize the Big-M method to convert Eq. (4) to the following linear constraints \cite{samani2019tri}:
\begin{align}
&  && \eta_t = b_{t}^1 \times (\pi_t - \lambda_t), &\forall t \in T \ \! \\ 
&  && x \geq \lambda_t - M \times (1-b_{t}^1), &\forall t \in T \ \! \\
&  && x \leq \lambda_t  + M \times b_{t}^1, &\forall t \in T  &&
\end{align} \label{eq:ax_2}
%
%


\noindent where $M$ is a large fixed parameter in the Big-M method. Next, equations (6)-(7) are transformed to the following:
\begin{align}
&  && P_t = \eta_t \times b_{t}^2 &\forall t \in T \\ 
&  && L_t = \eta_t \times (1-b_{t}^2), &\forall t \in T \\ 
&  && \eta_t \geq -M \times (1-b_{t}^2), &\forall t \in T \\ 
&  && \eta_t \leq M \times b_{t}^2, &\forall t \in T &&
\end{align} \label{eq:ax_3}
\noindent where $b^1$ and $b^2$ are binary variables.
By replacing $R_{t}$ in equations (15)-(18) with equation (12), the new MILP maximization problem will be formulated as follows:
\begin{align}
& {\text{maximize}} &&  Obj : \sum\limits_{t = 1}^T b_{t}^1 \times (\pi_t - \lambda_t)\\
& \text{subject to} && {\lambda}_t^{*} - m  \geq \lambda_t - M (1-b_{t}^1), &\forall t \in T \\
&  && {\lambda}_t^{*} - m  \leq \lambda_t + M \times b_{t}^1, &\forall t \in T \\
&  && b_{t}^1 \times (\pi_t - \lambda_t) \geq -M (1-b_{t}^2), &\forall t \in T \\ 
&  && b_{t}^1 \times (\pi_t - \lambda_t) \leq M \times b_{t}^2, &\forall t \in T \\
& && \sum\limits_{t = 1}^T (z_t - b_{t}^1) \times (\pi_t - \lambda_t) \leq \nonumber \\ 
& && \qquad \epsilon \times \sum\limits_{t = 1}^T z_t \times (\pi_t - \lambda_t) \\ 
&  &&  m^{min} \leq m \leq m^{max} \\
&  && z_t \leq b_t^1, &\forall t \in T  \\ 
&  && z_t \leq b_t^2, &\forall t \in T  \\
&  && z_t \geq b_t^1 + b_t^2 -1, &\forall t \in T \\
&  &&  0 \leq z_t \leq 1, &\forall t \in T &
\end{align} \label{eq:cnvx}

\color{black}
The optimization problem in (19)-(29) is the linearized version of the optimization problem in (3)-(9). The process of linearizing this optimization problem is done through (12)-(18). Note that, $z$ is a \emph{new} continuous auxiliary variable which takes the value of the multiplication of $b^1$ and $b^2$ to avoid the nonlinearity. Constraints (26)-(29) are added to the linearized optimization problem in order to create the required conditions for $z$ to be able to work as the multiplication of $b^1$ and $b^2$.
This final MILP optimization problem in (19)-(29) can be solved by using various commercial solvers.
\color{black}

\subsection{Step 2: Dynamic Node Labeling} 
Algorithm 1 is developed to dynamically \emph{label} each node for the next day, using the optimization-based results in Step 1.
For each node, first, we solve the negative and the positive spike capturing problems.
If the \emph{optimal objective value} for the negative spike capturing problem is greater than a threshold, then the node is labeled as Demand CB Node.
If the \emph{optimal objective value} for the positive spike capturing problem is greater than a threshold, then the node is labeled as Supply CB Node.
The two optimization problems are independent; hence, a node can be labeled \emph{both} as Demand CB Node and Supply CB Node. A node may also be labeled as No CB Node.

As another output of the optimization problem in (19)-(29), $m$ is used for generating the optimal price bid for each hour of the next day based on the label of each node.
If a node is labeled as Demand CB Node, Supply CB Node, or Both, then it will be considered for the next (final) step in the proposed convergence bidding strategy, as we will explain next.

\subsection{Step 3: Strategy Selection} \label{sec:propos}
In this section, we put together all the components, including the optimization-based price spike capturing method in Step 1 and the dynamic node labeling method in Step 2 to develop a new composite convergence bidding strategy. The new CB strategy makes use of each of three reverse engineering CB strategies based on their advantages and disadvantages. 

\begin{algorithm}[t]
	\caption{Dynamic Node Labeling}
	\label{algo1}
	\begin{algorithmic}[1]
	    \State \textbf{Input:} Outputs of Optimization-Based Spike Capturing
	    \State \textbf{Output:} Label and Optimal Price Bid for each Node
		\For {$n$ in Nodes}
    		\State Solve the negative spike capturing problem in (3)-(9).
    		\If {$Obj > \theta $}
        		\State Label $n$ as a \emph{Demand CB Node}
        		\State Optimal Price Bid $=\lambda_t^{*} - m$ 
    		\EndIf
    		\State Solve the positive spike capturing problem in (3), (6)-(11).
    		\If {$Obj > \theta $}
        		\State Label $n$ as a \emph{Supply CB Node}
        		\State Optimal Price Bid $=\lambda_t^{*} + m$ 
    		\EndIf    		

		\EndFor
	\end{algorithmic} 
\end{algorithm}

The inputs for this comprehensive strategy are the historical LMPs and the forecasted LMPs for the next day. The output is the type of strategy that should be used at each node for each hour of the next day.
The outline of this strategy is shown in Fig. \ref{fig:flowchrt}. Here, $a^{\lambda}$, $a^{\pi}$, and $a^{\delta}$ are the accuracy for the forecasted D-LMP, R-LMP, and the sign of the difference between D-LMP and R-LMP, respectively. 
As we can see, the first two strategies in this algorithm are based on the \emph{forecast accuracy} of the next day LMPs.
On the other hand, the third strategy does not use any price forecast data and instead, relies on the historical LMPs, optimization-based price spike capturing, and node labeling.
It must be mentioned that these three strategies are based on the three identified clusters of strategies in Section III.B.
\color{black}
The choice of the thresholds for the mentioned forecast accuracy defines the risk preference for the use of each strategy. Higher thresholds in Fig. \ref{fig:flowchrt}, mean less risk-seeking; and lower thresholds mean more risk-seeking.
\color{black}
The application of each strategy is as follows:
\begin{figure}[t]
 \centering
{\scalebox{0.66}{\includegraphics*{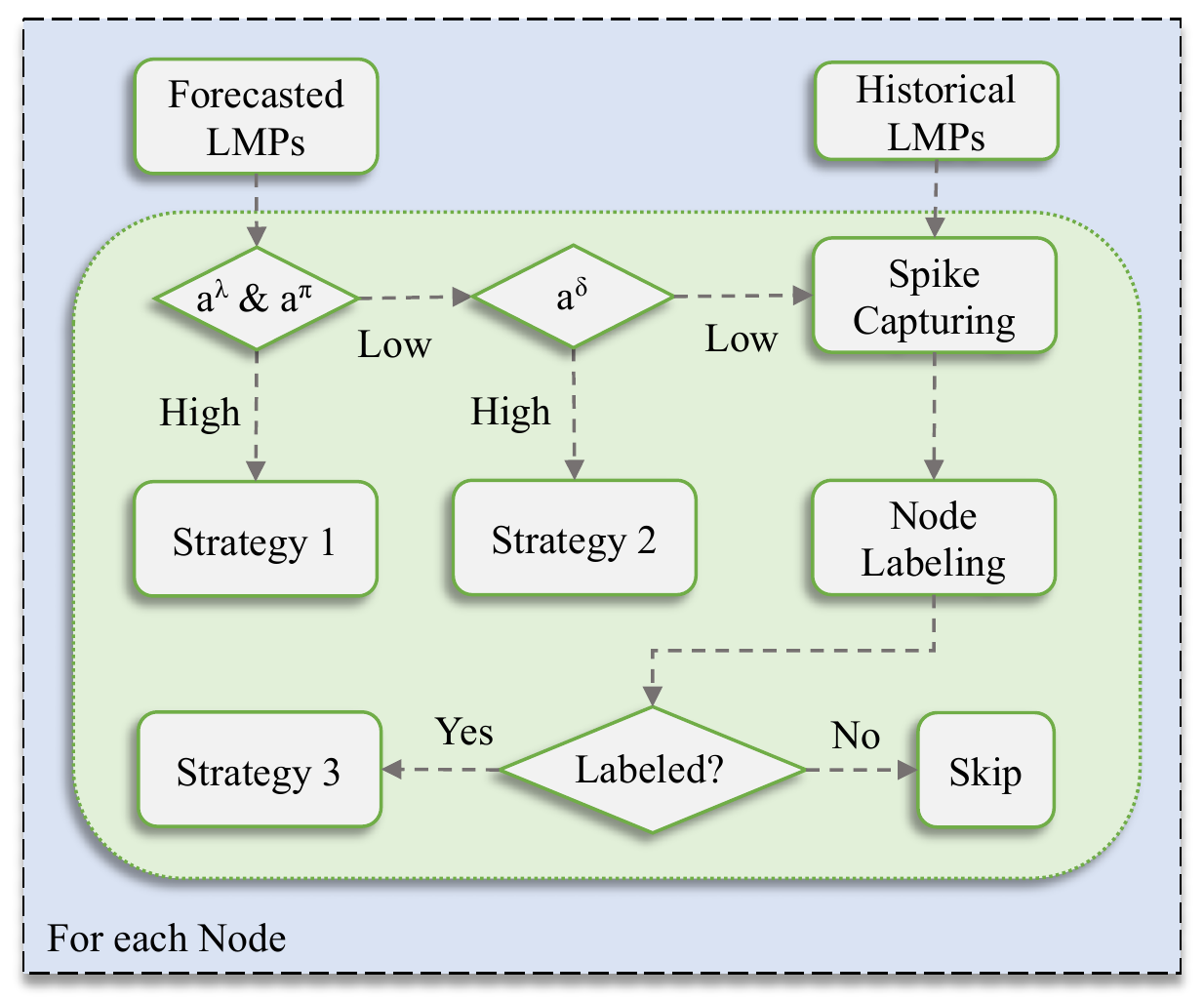}}}
\caption{The proposed comprehensive convergence bidding strategy for each hour of the next day and at each node in the market, based on the reversed engineered strategies of market participants.}
\label{fig:flowchrt}
\end{figure}

   \vspace{0.1cm}
   
\textbf{Selecting CB Strategy 1}: 
This strategy is selected if both D-LMP and R-LMP forecasts are available and have high accuracy. For each hour of the next day, if the forecasted D-LMP is higher than R-LMP, then a supply CB is submitted. If the forecasted R-LMP is higher than D-LMP, then a demand CB is submitted.
For a supply CB, the price components should be \emph{less} than the forecasted D-LMP but \emph{close} to it in order to avoid entering the market when D-LMP is unexpectedly low. Also, for a demand CB, the price components should be \emph{more} than the forecasted D-LMP but \emph{close} to it in order to avoid entering the market when D-LMP is unexpectedly high.

  \vspace{0.1cm}
  
\textbf{Selecting CB Strategy 2}: 
This strategy should be used if only the forecast for the sign of the difference between D-LMP and R-LMP is available and it has high accuracy.
For each hour of the next day, if the difference between D-LMP and R-LMP is \emph{positive}, then a supply CB is submitted; and if it is \emph{negative}, then a demand CB is submitted.
The price components for a supply CB must be \emph{much lower} than the average D-LMP; and for a demand CB, it must be \emph{much higher} than the average D-LMP on that hour in order to always be cleared in the market.
 
  \vspace{0.1cm}
  
\textbf{Selecting CB Strategy 3}:  
This strategy is used when accurate forecasting for LMPs are \emph{not} available at a node.
Using the historical LMPs in the spike capturing optimization problem and the node labeling algorithm, the optimal price bid and the type of CB are determined for a node.
As explained in Step 1, the optimal price component of the submitted CB is the best price bid to capture the price spikes; which leads to a situation where the convergence bidders participate in the CB market only occasionally, but when their CB is cleared they gain considerable profit.
This behavior is indeed justified, because in the absence of accurate price forecasting capability, one should avoid taking high risks.

\color{black}
It must be emphasized that, each strategy has its own importance and it forms part of the proposed enhanced composite bidding strategy. For example, although we showed that the performance of the Self-Scheduling Strategy is not as good as the other two strategies in the California ISO market, the Self-Scheduling Strategy does bring value to our composite bidding strategy when there is an accurate forecast only for the sign of the difference between D-LMP and R-LMP. If there is an accurate forecast for both D-LMP and R-LMP, then the Price-Forecasting Strategy would be a better choice.

\color{black}
Importantly, the construction of the above proposed algorithm also addresses Research Question 6. Here, the seemingly unprofitable (or low profitable) CB strategy is the second strategy. It is now incorporated as one part of an enhanced and profitable new bidding strategy.

Another note to mention is that, the overall architecture of our proposed composite bidding strategy is \emph{not} sensitive to or even directly related to the specific building of the explained clusters. That is, if other major clusters of bidding strategies emerge in the future, they too can potentially be incorporated into the architecture of our proposed composite bidding strategy by reveres engineering their main characteristics.

\color{black}

Before ending this section, it should be mentioned that the quantity of submitted CBs (MWh) at each hour, depends on the available credit for each market participant with the California ISO. Accordingly, a unit value is considered in the proposed bidding strategy for the quantity of submitted CBs, which is aligned with other studies in the literature such as \cite{baltaoglu2018algorithmic}.


\subsection{Case Study}
In this section, we analyze the performance of the proposed comprehensive convergence bidding strategy.
Since providing an accurate and realistic forecast for D-LMPs and R-LMPs is out of the scope of this work, here we assume that we have the same forecasting accuracy as Alias ID 9, which is the most lucrative CB market participant in the California ISO market during the period of this study.
Our goal here is to examine how Alias ID 9 could improve its performance in 2019, if it had used our proposed composite strategy.
\color{black}
The value for $M$ in the Big-M method is tuned to be a sufficiently large number. In this regard, it is set to 3000. Also, the boundaries for $m$ in equation (25) are set to 30 and 200, respectively; which are based on our observations on the real-world data.
\color{black}
The analyses in this section are done in Python and the optimization problems are solved by using Gurobi within the Pyomo package on a PC with Intel Xeon Silver 4208 CPU @2.10GHz and 128 GB RAM.

As mentioned before, Alias ID 9 mostly participated in the main Hubs and DLAPs by utilizing the Price-Forecasting Strategy.
For other nodes, we use one year of historical data before each day in 2019 and run the dynamic node labeling algorithm including the optimization-based spike capturing problem.
By adjusting the only two hyperparameters, $\epsilon$ and $\theta$, the nodes with the potential to use the Opportunistic Strategy are labeled and the optimal price bid is submitted for each hour of the next day.
The submitted CBs have one step and the quantity of each submitted CB is considered as the average of Alias ID 9's submitted CBs equal to 50 MW, instead of a unit value.
All the above assumptions match the overview of the actual market data in Table \ref{tab2}.

\begin {table}[t]
\centering
\caption {Results for Three Analyzed Cases Based on \\ Different Values of the Hyperparameters.}
\label{tab4}
\begin{center}
    \begin{tabular}{| c | c | c | c | c | c | c |}
    \hline
Case & $\epsilon$ & $\theta$ & Node & Day & $\eta$ (\$M) & LPR (\%) \\ \hline
1    & 0.01  & 100    & 148  & 227 & 3.89             & 24.84          \\ \hline
2    & 0.001 & 1000   & 32   & 66  & 2.13             & 11.14            \\ \hline
3    & 0.0001  & 2000    & 25  & 40 & 1.55             & 5.25          \\ \hline
\end{tabular}
\end{center}
\end{table}

\begin{figure}[t]
 \centering
{\scalebox{0.45}{\includegraphics*{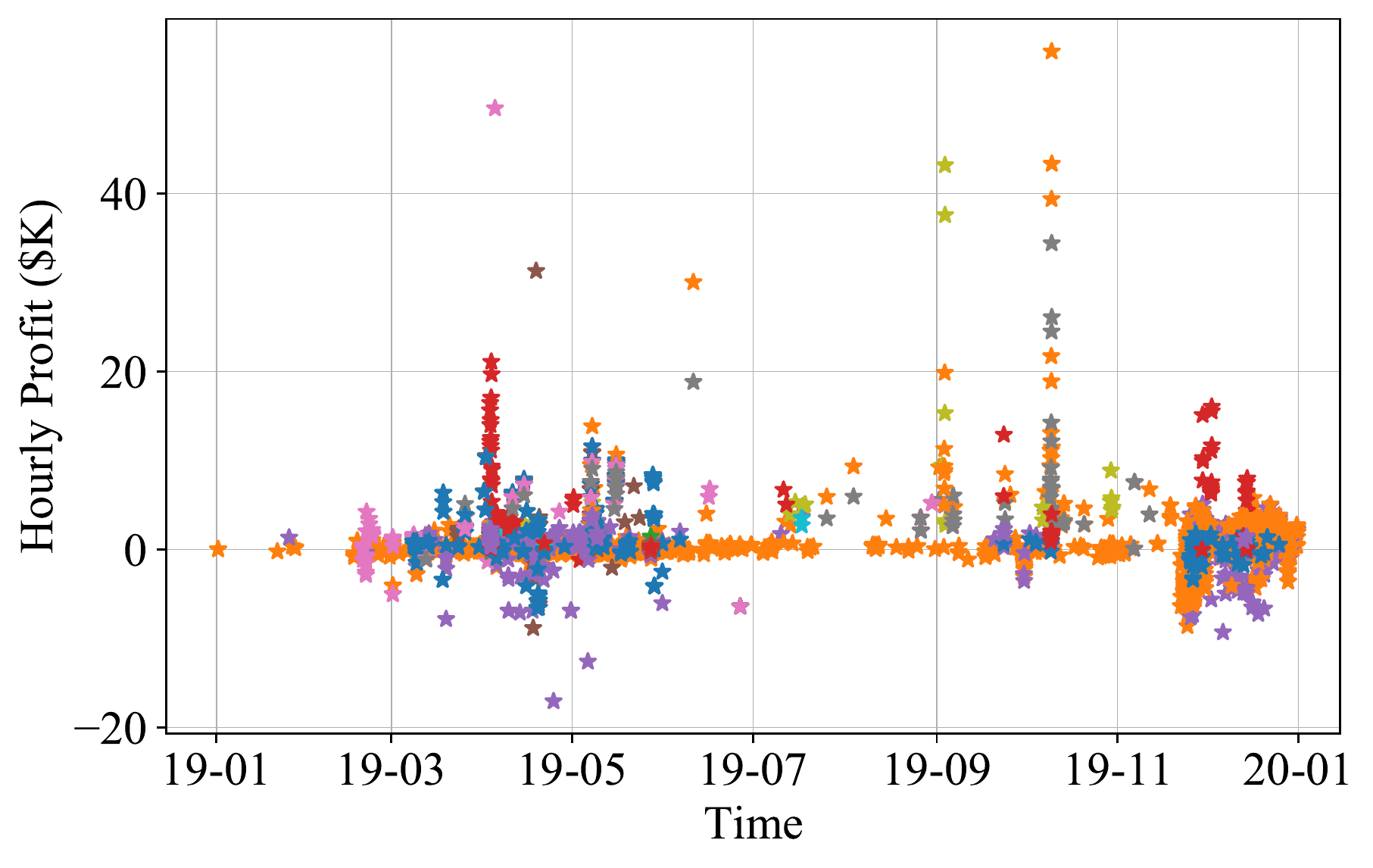}}}
\caption{The \emph{additional} hourly net profit in Case 1. Each color represents one of the 148 labeled nodes where the submitted CBs are cleared.}
\label{fig:add_prft_cs2}
\end{figure}

\begin{figure}[t]
 \centering
{\scalebox{0.45}{\includegraphics*{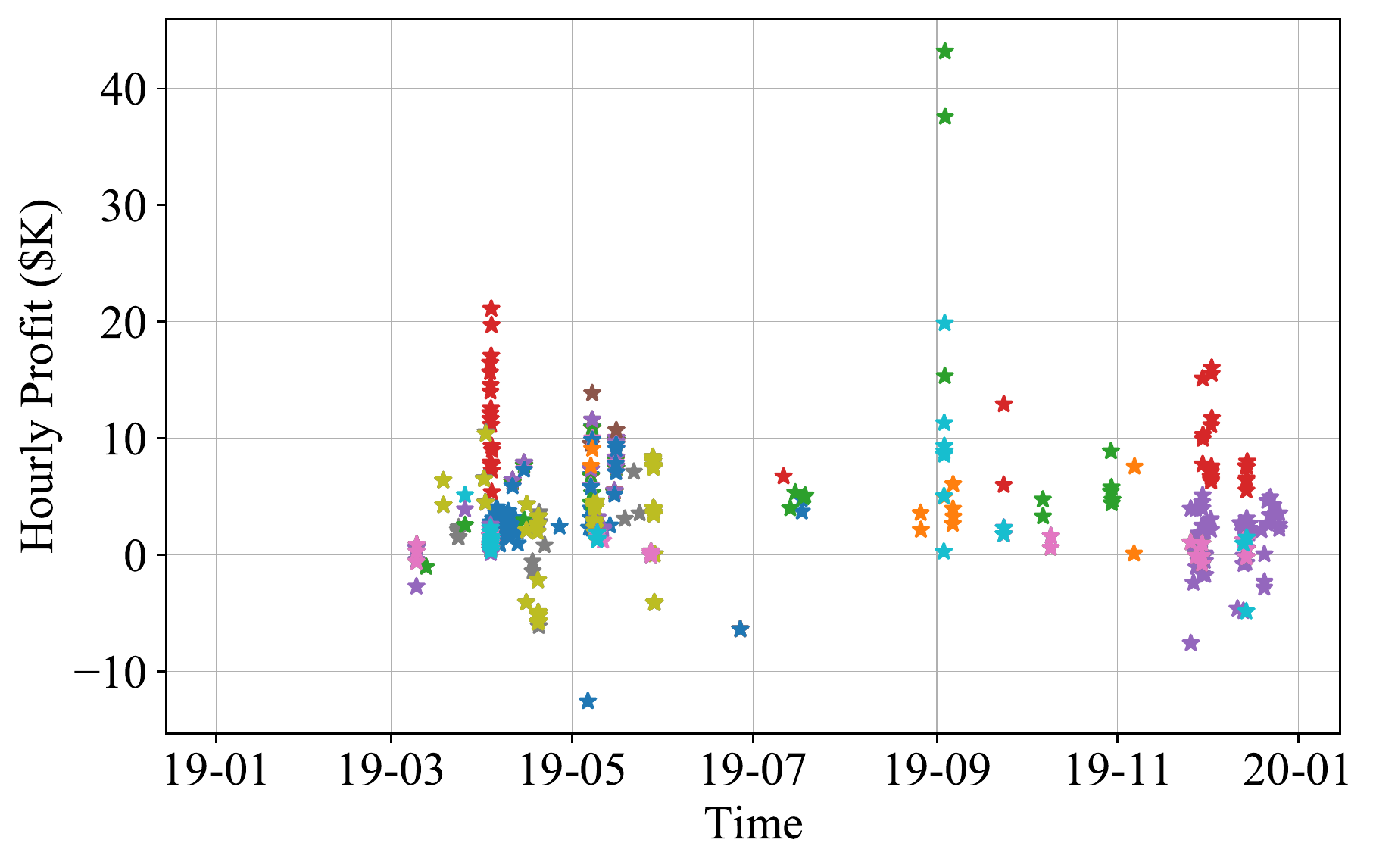}}}
\caption{The \emph{additional} hourly net profit in Case 2. Each color represents one of the 32 labeled nodes where the submitted CBs are cleared.}
\label{fig:add_prft}
\end{figure}

Table IV shows three cases based on different values for the hyperparameters. We can see that, by tightening the constraints (decreasing $\epsilon$ and increasing $\theta$) from Case 1 to Case 3, the number of nodes and days that the submitted CBs are cleared has decreased.
\color{black}
As we defined in (2), LPR indicates the level of loss compared to the level of profit. Also, recall from Section IV.A that $\eta$ denotes the total net profit.
\color{black}
As we can see in Table IV, by tightening the constraints, although the total net profit is decreased, the loss-to-profit-ratio as an important factor in the Opportunistic Strategy is also decreased.
\color{black}
Therefore, these two parameters can be used as control knobs by the market participant to suitably adjust the level of risk seeking in this composite bidding strategy.
\color{black}
Figs. \ref{fig:add_prft_cs2} and \ref{fig:add_prft} show the additional hourly net profit for Case 1 and Case 2. 
\color{black}
Note that, notation \$K means \$1,000.
\color{black}
In Fig. \ref{fig:add_prft}, for Case 2 as a moderate case, it is shown that by submitting CBs at 32 nodes and entering the CB market in only 66 days, Alias ID 9 could earn an \emph{additional} net profit of \$2.13 million. This is a 43\% increase in Alias ID 9's net profit in 2019, compared to its current net profit of \$4.9 million.
\color{black}
In order to further extended the assessment of the increase in net profit, Table V shows the original annual net profit of four market participants (Alias IDs) in 2019 and compares them with the corresponding annual net profit of the same market participant in case she had used the proposed convergence bidding method in Algorithm 1. Similar to the previous test case for Alias ID 9 as the most lucrative market participant, here we use the average quantity in MWh from Table II for the size of the CBs for each market participant. We can see that all these four market participants that were mainly focused on the third strategy based on Fig. 10, could have significantly benefited from the proposed convergence bidding strategy.
\color{black}
\color{black}

\begin {table}[t]
\color{black}
\centering
\caption {Comparing the Performance of the Proposed Method with \\ Four Alias ID in 2019 which Mostly Used the Third Strategy.}
\label{tab5}
\begin{center}
  \begin{tabular}{| c | c | c | c |}
  \hline
Alias & Original & New & Improvement  \\ 
ID & Net Profit (\$M)  & Net Profit (\$M)   & (\$M)  \\ \hline
1    &  0.30 &	1.27	& 0.97 \\ \hline
5    &  0.31 &	3.26	& 2.95  \\ \hline
6    &  0.44 &	2.1	&  1.66 \\ \hline
8    &  0.58 &	0.84	& 0.26  \\ \hline
\end{tabular}
\end{center}
\end{table}
\color{black}

\section{Conclusions and Future Work} \label{sec:conclusions}
This paper provided a data-driven analysis of real-world electricity market data from the California ISO market to \emph{understand}, \emph{reverse engineer}, and \emph{enhance} the behavior of convergence bidders.
It was discussed that a total of 20 CB market participants currently have a considerable presence in the California ISO that accounts for 72\% to 84\% of the entire CB market.
The different bidding characteristics of these most present market participants were analyzed.
Next, four quantitative features were extracted from all the submitted CBs; and by using the HDBSCAN algorithm, three main clusters of CB strategies were identified.
The characteristics and the performance of each identified cluster of strategies were analyzed and some of their \emph{advantages} and \emph{disadvantages} were investigated.

Two interesting discoveries were discussed. First, the Opportunistic Strategy does \emph{not} match any of the convergence bidding strategies that currently exist in the literature.
Second, most papers in the literature are focused on the Self-Scheduling Strategy, while in practice, this strategy is \emph{less} common among the market participants in the California ISO. 

After reverse engineering the convergence bidding strategies of the real-world market participants in the California ISO market, a new comprehensive \emph{composite} CB strategy was proposed. It was shown that the proposed strategy, optimally utilizes the advantages of various identified reverse engineered strategies under different market conditions. The new strategy was developed in three steps:
First, by focusing on the Opportunistic Strategy as the newly discovered strategy, an optimization-based algorithm was proposed to maximize the total net profit of the market participant by capturing the price spikes.
Second, an algorithm was developed to dynamically label each node based on the solution of the optimization problem in the first step.
Third, by using the results from the first two steps for the Opportunistic Strategy, as well as by combining them with the Price-Forecasting Strategy and the Self-Scheduling Strategy, a strategy selection algorithm was proposed to complete a comprehensive composite CB strategy.
It was shown in a case study that the annual profit of the most lucrative market participant could increase by over 40\% if the proposed comprehensive strategy had been used. 

The study in this paper can be extended in different directions.
For example, we may investigate how the identified real-world CB strategies may positively or negatively affect price convergence and the efficiency of the electricity market.
In other words, we may investigate the system-level impact of the identified real-world bidding strategy and/or the proposed composite bidding strategy. 
We may also analyze and reverse engineer the bidding strategy of the market participants that also submit physical bids.
\color{black}
Another interesting path for potential future research is to investigate the impact of the behavior of CB market participants on each other; i.e., by using concepts and methods in Game Theory.
\color{black}

\vspace{0.2cm}

\bibliographystyle{IEEEtran}
\bibliography{References.bib}

\end{document}